\documentclass{article}

\usepackage{arxiv}

\usepackage[utf8]{inputenc} 
\usepackage[T1]{fontenc}    
\usepackage{hyperref}       
\usepackage{url}            
\usepackage{booktabs}       
\usepackage{amsfonts}       
\usepackage{nicefrac}       
\usepackage{microtype}      
\usepackage{lipsum}		
\usepackage{caption}
\usepackage{subcaption}
\usepackage{graphicx}
\usepackage{rotating}
\usepackage{threeparttable}
\usepackage{lscape}
\usepackage[authoryear,round,sort]{natbib}
\setcitestyle{aysep={}}
\usepackage{doi}
\usepackage{amsmath}
\usepackage[shortlabels]{enumitem}
\usepackage{multirow}
\usepackage{changepage}

\title{Comparative Analysis of Weather-Based Indexes and the Actuaries Climate Index\textsuperscript{TM} for Crop Yield Prediction and Weather-Derivative Pricing}

\author{ \href{https://orcid.org/0000-0002-9455-6397}{\includegraphics[scale=0.06]{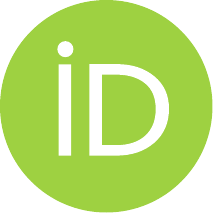}\hspace{1mm}Cem~Yavrum} \\ 
	Institute of Applied Mathematics\\
	Middle East Technical University\\
	06800, Ankara, Türkiye \\
	\texttt{cyavrum@metu.edu.tr} \\
	\And
	\href{https://orcid.org/0000-0001-5647-7973}{\includegraphics[scale=0.06]{orcid.pdf}\hspace{1mm}A. Sevtap~Selcuk-Kestel} \\
	Institute of Applied Mathematics\\
	Middle East Technical University\\
	06800, Ankara, Türkiye \\
	\texttt{skestel@metu.edu.tr} \\
	\And
	\href{https://orcid.org/0000-0002-2016-7524}{\includegraphics[scale=0.06]{orcid.pdf}\hspace{1mm}Jos\'e~Garrido} \\
	Concordia University\\
	Montreal, Canada\\
	\texttt{jose.garrido@concordia.ca} \\
}

\renewcommand{\shorttitle}{Climate Index Applications in Yield Prediction and Derivative Pricing}

\hypersetup{
pdftitle={A template for the arxiv style},
pdfsubject={q-bio.NC, q-bio.QM},
pdfauthor={David S.~Hippocampus, Elias D.~Striatum},
pdfkeywords={First keyword, Second keyword, More},
}

\graphicspath{ {figures/} }

\begin{document}
\maketitle

\begin{abstract}
	Climate change poses significant challenges to the agricultural and financial sectors, affecting crop productivity and the overall financial stability. This study evaluates the robustness of the Actuaries Climate Index\textsuperscript{TM} (ACI), a relatively recent tool to measure the impact of climate change, by comparing its explanatory power to well-established weather-based indexes (WBIs) across two key sectors. In the agricultural context, the yields of three major crops are predicted using generalized statistical models and advanced machine learning algorithms with climate indexes as explanatory variables. To enhance model reliability and address multicollinearity among weather-related variables, the study also incorporates both principal component analysis and functional principal component analysis. A total of 22 models, each constructed with different sets of explanatory variables, illustrate the significant impact of wind speed and sea-level changes, alongside temperature and precipitation, on crop yield variability across six regions of the United States. For the financial market application, the analysis adapts the weather-derivative framework, as it is a critical instrument for energy companies, insurers, and agribusinesses seeking to hedge against weather-related risks. By analyzing the payoffs of derivative contracts that use WBIs and ACI components as underlying variables, the findings reveal that the ACI framework holds a strong potential as a comprehensive climate risk indicator, not only for the agricultural sector but also for the finance and insurance industries.
\end{abstract}


\section{Introduction}
Climate change refers to long-term shifts in temperature and weather. Recent reports from the Intergovernmental Panel on Climate Change highlight the significant increase in global temperatures over the past few decades \citep{r24, r25}. The consequences of this warming extend across multiple domains, adversely affecting human health, global food security, financial stability, and the insurance industry. From a public health perspective, \citet{r26} projects that climate change could lead to an additional 250,000 deaths per year between 2030 and 2050. In the financial sector, \citet{r29} report that direct losses from natural hazards in the United States (U.S.) are steadily increasing and may reach between \$300 and \$400 billion within a single decade, driven by the increasing frequency and severity of extreme weather events. In terms of global food security, empirical studies by \citet{r32}, \citet{r33}, and \citet{r35} demonstrate that the yields of major crops are significantly affected negatively by increasing temperatures and changing precipitation patterns. Similarly, within the insurance industry, \citet{r31} discuss the profound impact of climate change across various insurance sectors, underscoring the urgency for standardized methodologies to measure and quantify climate risks.

Climate impact is mostly measured and interpreted by indexes. In the literature, various indexes such as the Climate Extremes Index \citep{r36}, Greenhouse Climate Response Index \citep{r36}, Palmer Drought Severity Index \citep{r37}, Cooling Degree Days \citep{r1}, Heating Degree Days \citep{r1} and Cumulative Rainfall Index \citep{r2} are developed to quantify climate change, enabling researchers and policymakers to assess systematically its impact. In recent years, the Actuaries Climate Index\textsuperscript{TM} (ACI) has emerged as a pivotal tool for understanding and monitoring climate risks, offering valuable insights for insurance companies, governments, and the general public. Developed in 2016 by the four major North American actuarial societies, the \citet{r3} uses historical meteorological data from the U.S. and Canada to track extreme conditions in air temperature, precipitation, drought, wind, and sea-level rise.

The practical methodology of the ACI has quickly been adapted for use in other countries. \citet{r38} demonstrates that the ACI framework is effectively applicable to regions like the United Kingdom and Europe. The Australian Actuaries Climate Index \citep{r39}, which is tailored to Australian meteorological conditions, is introduced, while \citet{r40} develop the Turkish Actuaries Climate Index, specifically for the Ankara region in Türkiye. Similarly, \citet{r4} launch the Iberian Actuarial Climate Index, encompassing Spain and Portugal, marking the first application of the ACI framework to the European continent. Most recently, \citet{r41} and \citet{r71} extend this framework to France and Italy, respectively, creating the French and Italian Actuarial Climate Indexes. The continued international adoption of the ACI methodology demonstrates a growing global interest to quantifying, monitoring, and managing climate risks through standardized and comparable metrics.

Several studies in the literature also demonstrate the applicability of the ACI in various fields, particularly in insurance and finance. \citet{r42} analyze ACI's predictive power for stock returns of agriculture-related companies in the U.S. and Canada, concluding that ACI could support a profitable trading strategy. \citet{r23} assess ACI's effectiveness in predicting corn yields in the Midwest region of the U.S. using linear and probit regression models, finding satisfactory predictive power. \citet{r43} explore the relationship between the Spanish Actuarial Climate Index and hailstorm-related insurance losses for grapes grown for wine producing, through linear and quantile regression models.

However, given its relatively recent addition, there remains a notable gap in the literature where a systematic evaluation of the ACI’s predictive capability against traditional weather indexes would provide a significant contribution, thereby enhancing its applicability across broader fields and sectors. By comparing the explanatory power of the ACI with that of well-established weather-based indexes (WBIs) commonly used in the energy and derivatives markets, this study aims to evaluate the robustness of the ACI as a tool for measuring climate impacts in both the agricultural sector and financial markets. In the agricultural context, weather-based indexes are used to predict the yields of three major crops using regression models, namely Generalized Linear Model (GLM) and Generalized Additive Model (GAM), alongside advanced machine learning algorithms such as Extreme Gradient Boosting (XGB) and Light Gradient Boosting Machine (LGBM). For the financial market application, the analysis focuses on the payoffs of weather derivative contracts, specifically uncapped call options that use WBIs and ACI components as underlying variables. To the best of our knowledge, beyond comparing the ACI with traditional weather indexes, this study makes an additional novel contribution by presenting the first application of the ACI within a weather-derivative framework. It shows how the ACI could become a critical instrument for energy companies, insurers, and agribusinesses seeking effective hedging mechanisms against weather-related risks. In this context, the study also provides one of the first comprehensive evaluations of the ACI across both crop yield prediction and weather derivative pricing frameworks.

Regarding crop yield prediction, this study provides a systematic evaluation of the individual and joint contributions of weather indexes by fitting 22 distinct models, each constructed using different combinations of underlying weather measures and the corresponding ACI components. This structured framework enables a comprehensive assessment of how each index and their interactions influence model performance. Since all variables are weather-based, they may share common weather driven patterns, potentially leading to multicollinearity issues in regression models. Additionally, achieving comparable model performance with fewer variables is both practical and desirable. To address these challenges, dimension reduction techniques specifically, Principal Component Analysis (PCA) and Functional Principal Component Analysis (FPCA) are applied to identify the most influential variables in the data sets. This approach not only mitigates multicollinearity issues but also simplifies the models, improving efficiency and interpretability.

Building on the proposed framework, the study further investigates the extent to which ACI components and traditional weather-based indexes provide consistent and reliable representations of climate-related impacts across different application domains. In particular, it explores the potential role of additional weather variables-such as drought, wind, and sea-level anomalies-in improving model performance beyond conventional temperature and precipitation-based measures. The analysis also explores whether consistency in derivative payoffs supports the use of the ACI as a comprehensive climate risk indicator, with potential applicability in both agricultural and financial contexts.

The paper is organized as follows: Section \ref{sec2} summarizes the climate indexes, whereas Section \ref{sec3} describes the proposed methodology for predicting crop yields and pricing weather derivatives based on these indexes. The properties of the data and the results of the analyses are provided in Section \ref{sec4}. Section \ref{sec44} discusses the main findings by providing a detailed interpretation of the results and their broader implications. Finally, Section \ref{sec5} concludes the paper by summarizing the key findings. The detailed structure of the explanatory variables across each model, the descriptive statistics of each index, and the comprehensive performance results for each method are reported in the Appendix.

\section{Climate Indexes} \label{sec2}

We focus on weather-based indexes, specifically Cooling Degree Days (CDD), Heating Degree Days (HDD), and the Cumulative Rainfall Index (PRE), which are directly linked to temperature and precipitation. These indexes hold particular advantages because they rely on a single weather variable, making them not only simple to interpret but also straightforward to compute. Their relative ease of use makes them especially suitable for comparative analyses with the components of the Actuaries Climate Index\textsuperscript{TM}.

This section is structured as follows. First, Section \ref{2.1.} introduces the definitions and formulations of the weather-based indexes. Section \ref{2.2.} then presents the Actuaries Climate Index\textsuperscript{TM} (ACI) and its components. Finally, Section \ref{2.3.} introduces a structured grouping of these indexes based on the weather conditions they represent, providing a consistent framework for the comparative analysis conducted in the empirical section.

\subsection{Weather-Based Indexes}\label{2.1.}

CDD and HDD are temperature-based indexes, with CDD capturing warm weather conditions and HDD representing cool weather conditions. PRE, on the other hand, quantifies rainfall amounts. These indexes are initially calculated on a daily basis, and their values for longer intervals, such as monthly or meteorological seasons, are derived by summing the daily data. These indexes capture different and complementary aspects of weather conditions and therefore do not overlap, but rather represent distinct dimensions of climatic variability. Briefly, CDD measures how much the daily average outside air temperature exceeds a base level temperature, and HDD measures the opposite, the temperature falls below a base level. The base temperature of 65°F (18°C) is widely adopted in the literature, as it represents a standard benchmark for heating and cooling demand \citep{r1, r74, r75}. Since this study focuses on U.S. data, the use of 65°F ensures consistency with both the existing literature and industry practice in weather derivatives.

Let $T_i^{\text{max}}$ and $T_i^{\text{min}}$ denote the maximum and minimum temperatures measured on day $i$. The average temperature for day $i$ is defined as:

\begin{equation}\label{eq1}
	T_i = \frac{T_i^{\text{max}} + T_i^{\text{min}}}{2}.
\end{equation}

If the average temperature is higher than 65°F for CDD and lower than 65°F for HDD, the corresponding daily values are defined as:

\begin{equation}\label{eq2}
	CDD_i = max\{T_i - 65, 0\},
\end{equation}
\begin{equation}\label{eq3}
	HDD_i = max\{65 - T_i, 0\},
\end{equation}
where $T_i$ denotes the average temperature on day $i$. Additionally, PRE represents the total volume of rainfall accumulated over a specified time period and is calculated by summing all daily rainfall measurements \citep{r2}:

\begin{equation}\label{eq4}
	PRE = \sum_{i=1}^{L} r_i,
\end{equation} 
where $r_i$ denotes the daily total rainfall amount recorded on day $i$, measured in millimeters (mm), and $L$ is the total number of days in the chosen time frame, such as a month or a meteorological season.

\subsection{Actuaries Climate Index\textsuperscript{TM}}\label{2.2.}

The ACI is designed to improve understanding of climate trends and their potential impact, incorporating various climate variables that signify extreme weather conditions. It comprises six components: high temperature (T90), low temperature (T10), precipitation (P), drought (D), wind speed (W), and sea-level (S). Since each component reflects different weather conditions, the ACI uses standardized values (anomalies) for all components to integrate them into a single index. Standardization is based on the mean and standard deviation of each component during the designated reference period of 1961 to 1990 \citep{r3}. The components of the ACI are defined as follows:

\begin{itemize}\itemsep0em
	\item T90 represents the percentage of days in a month when daily temperatures exceed the 90th percentile of the corresponding days in the reference period.
	\item T10 reflects the percentage of days when daily temperatures fall below the 10th percentile of the reference period.
	\item P, denoted as Rx5day, measures the maximum amount of rainfall recorded over any five consecutive days within a month.
	\item D represents the annual maximum number of consecutive dry days during which daily precipitation is below 1 millimeter.
	\item W quantifies the monthly frequency of days on which the daily mean wind power exceeds the 90th percentile of the reference period.
	\item S captures the combined impact on coastal shorelines resulting from long-term sea-level rise.
\end{itemize}

The standardized values of each component (anomalies) are computed both on a monthly and seasonal basis. For each month ($j$) and year ($q$), they are expressed as \citep{r4}:

\begin{equation}\label{eq5}
	T90_{STD}(j,q)  = \frac{T90(j,q) - \mu_{T90(j)}}{\sigma_{T90(j)}},
\end{equation}

\begin{equation}\label{eq6}
	T10_{STD}(j,q)  = \frac{T10(j,q) - \mu_{T10(j)}}{\sigma_{T10(j)}},
\end{equation}

\begin{equation}\label{eq7}
	P_{STD}(j,q)  = \frac{Rx5day(j,q) - \mu_{Rx5day(j)}}{\sigma_{Rx5day(j)}},
\end{equation}

\begin{equation}\label{eq88}
	D_{STD}(j,q)  = \frac{D(j,q) - \mu_{D(j)}}{\sigma_{D(j)}},
\end{equation}

\begin{equation}\label{eq99}
	W_{STD}(j,q)  = \frac{W(j,q) - \mu_{W(j)}}{\sigma_{W(j)}},
\end{equation}

\begin{equation}\label{eq89}
	S_{STD}(j,q)  = \frac{S(j,q) - \mu_{S(j)}}{\sigma_{S(j)}}.
\end{equation}

Here, $\mu_{\cdot(\cdot)}$ and $\sigma_{\cdot(\cdot)}$ are the mean and standard deviation of corresponding components for the reference period, respectively. The composite ACI is derived by averaging the standardized values of all six components, with the T10 component being assigned a negative value, and expressed as \citep{r3}:

\begin{equation}
	ACI = \frac{T90_{STD} - T10_{STD} + P_{STD} + D_{STD} + W_{STD} + S_{STD}}{6}.
\end{equation}

The negative sign associated with the T10 component reflects its inverse relationship with warming conditions. Specifically, a decrease in cold extremes (i.e., lower values of T10) corresponds to an increase in overall warming. To ensure that all components consistently contribute in the same direction to the index, the sign of the T10 component is reversed. For further details, the reader is referred to \cite{r3}.

\subsection{Index Grouping}\label{2.3.}

The weather-based indexes (CDD, HDD, and PRE) and the components (T90, T10, and P) of the ACI are conceptually related in that they capture similar underlying climate phenomena, but differ in their construction and interpretation. Specifically, CDD and HDD are based on cumulative temperature deviations relative to a fixed threshold, whereas the ACI temperature components (T90 and T10) measure the frequency of extreme temperature events relative to a historical reference period. Similarly, PRE represents cumulative rainfall amounts, while the precipitation component (P) of the ACI captures extreme precipitation events. The grouping of indexes, established to enable meaningful and structured comparisons, is summarized in Table \ref{tab1}.

\clearpage

\begin{table}[hbt!]
	\centering
	\caption{Structure of groups}
	\label{tab1}
	\scalebox{0.8}{
		\begin{tabular}{|c|c|c|c|}
			\hline
			\textbf{Group} & \textbf{\begin{tabular}[c]{@{}c@{}}Weather-Based\\ Indexes (WBIs)\end{tabular}} & \textbf{\begin{tabular}[c]{@{}c@{}}Components \\ of the ACI\end{tabular}} &  \textbf{Measurement} \\ \hline
			\textbf{1}     & CDD                                                                     & T90                                                                          & Warm weather         \\ \hline
			\textbf{2}     & HDD                                                                     & T10                                                                          & Cool weather         \\ \hline
			\textbf{3}     & PRE                                                                     & P                                                                          & Precipitation        \\ \hline
	\end{tabular}}
\end{table}

While these indexes are not mathematically equivalent, they provide alternative representations of similar climate conditions. This distinction enables a meaningful comparison of traditional weather-based indexes and the ACI in terms of their explanatory power in the following agricultural and financial market applications.

\section{Proposed Methodology} \label{sec3}

Based on the main objectives of evaluating the robustness of the ACI as a tool for measuring climate impacts in both the agricultural and financial sectors, this section presents the proposed methodological framework. The analysis is organized around two main application areas: crop yield prediction and weather derivative pricing.

Section \ref{3.1.} focuses on the agricultural application and develops the modeling framework for crop yield prediction. It begins with a time-trend analysis (Section \ref{3.1.1.}) to address non-stationarity in the yield data. Next, the model specifications are introduced (Section \ref{3.1.2.}), including both regression-based and machine learning approaches. The evaluation procedure is then presented in Section \ref{3.1.3.}, where a time series cross-validation framework is used to assess model performance.

Section \ref{3.2.} presents the financial market application, where the use of climate indexes in weather derivative pricing is examined. Section \ref{3.2.1.} introduces the structure of the derivative contracts considered in the analysis, while Section \ref{3.2.2.} outlines the pricing methods used to evaluate their performance.

\subsection{Crop Yield Prediction}\label{3.1.}

The predictive capabilities of climate indexes on crop yields are evaluated using both traditional statistical models and modern machine learning algorithms. In the literature, regression models \citep{r47, r48, r49} and machine learning techniques \citep{r50, r51, r52, r57, r58, r59} are widely used for crop yield prediction, largely due to their ease of implementation, interpretability, and computational efficiency. In this study, the GLM \citep{r9} and GAM \citep{r10, r11}, which are favored for their flexibility in handling distributional assumptions, are used. To complement these approaches and highlight the advantages of machine learning methods over traditional statistical models, two widely used gradient boosting algorithms, XGB \citep{r45} and LGBM \citep{r46}, are also incorporated. 

Crop yields are influenced not only by weather conditions during the harvest period but also by extreme weather events occurring throughout the entire growing cycle, from planting to harvest. Adverse conditions such as temperature extremes and precipitation anomalies can have significant cumulative effects on crop development and productivity. To capture the full spectrum of potential impacts, the meteorological seasonal values of each index covering the entire year are included in the analysis. Additionally, the timing of harvest for each crop is carefully considered to ensure that the analysis encompasses the complete year leading up to harvest, thereby reflecting all relevant weather impacts. The analyses fit a total of 22 models, each incorporating explanatory variables representing different weather conditions and their combinations, as summarized in Table \ref{tab2} and further explained below. Additionally, a detailed representation of the structure of the explanatory variables included in each model is provided in Appendix \ref{tab22}. Comprehensive information on the data sets, along with the crop yields fitted by each model, is presented in Section \ref{sec4}.

\begin{table}[hbt!]
	\centering
	\caption{Explanatory variables included in predicting crop yields}
	\label{tab2}
	\scalebox{0.76}{
		\begin{tabular}{|cccc|c|c|}
			\hline
			\multicolumn{1}{|c|}{\textbf{\begin{tabular}[c]{@{}c@{}}Model\end{tabular}}} & \multicolumn{1}{c|}{\textbf{\begin{tabular}[c]{@{}c@{}}Explanatory\\      Variables\end{tabular}}} & \multicolumn{1}{c|}{\textbf{\begin{tabular}[c]{@{}c@{}}Model\end{tabular}}} & \textbf{\begin{tabular}[c]{@{}c@{}}Explanatory\\      Variables\end{tabular}} & \textbf{\begin{tabular}[c]{@{}c@{}}Model\end{tabular}} & \textbf{\begin{tabular}[c]{@{}c@{}}Explanatory\\      Variables\end{tabular}} \\ \hline
			\multicolumn{1}{|c|}{\textbf{1}}     & \multicolumn{1}{c|}{CDD}                                                                           & \multicolumn{1}{c|}{\textbf{8}}     & T90                                                                           & \textbf{15}    & T90-T10-P-W                                                                   \\ \hline
			\multicolumn{1}{|c|}{\textbf{2}}     & \multicolumn{1}{c|}{HDD}                                                                           & \multicolumn{1}{c|}{\textbf{9}}     & T10                                                                           & \textbf{16}    & T90-T10-P-D                                                                   \\ \hline
			\multicolumn{1}{|c|}{\textbf{3}}     & \multicolumn{1}{c|}{PRE}                                                                           & \multicolumn{1}{c|}{\textbf{10}}    & P                                                                             & \textbf{17}    & T90-T10-P-S                                                                   \\ \hline
			\multicolumn{1}{|c|}{\textbf{4}}     & \multicolumn{1}{c|}{CDD-HDD}                                                                       & \multicolumn{1}{c|}{\textbf{11}}    & T90-T10                                                                       & \textbf{18}    & T90-T10-P-W-D                                                                 \\ \hline
			\multicolumn{1}{|c|}{\textbf{5}}     & \multicolumn{1}{c|}{CDD-PRE}                                                                       & \multicolumn{1}{c|}{\textbf{12}}    & T90-P                                                                         & \textbf{19}    & T90-T10-P-W-S                                                                 \\ \hline
			\multicolumn{1}{|c|}{\textbf{6}}     & \multicolumn{1}{c|}{HDD-PRE}                                                                       & \multicolumn{1}{c|}{\textbf{13}}    & T10-P                                                                         & \textbf{20}    & T90-T10-P-D-S                                                                 \\ \hline
			\multicolumn{1}{|c|}{\textbf{7}}     & \multicolumn{1}{c|}{CDD-HDD-PRE}                                                                   & \multicolumn{1}{c|}{\textbf{14}}    & T90-T10-P                                                                     & \textbf{21}    & T90-T10-P-W-D-S                                                               \\ \hline
			\multicolumn{4}{|c|}{}                                                                                                                                                                                                                                          & \textbf{22}    & ACI                                                                           \\ \hline
	\end{tabular}}
\end{table}

The first seven models summarized in Table \ref{tab2} are built using only WBIs and their combinations, providing a baseline for comparison. Models 8-14 replace these with ACI components that correspond one-to-one with the WBIs, enabling direct performance comparisons between the two index types (see Table \ref{tab1}). To extend the analysis, additional ACI components (W, D, S) that are not represented in the first 14 models are sequentially incorporated into Models 15-21. Finally, Model 22 uses the composite ACI, which aggregates all six components into a single index, offering a comprehensive measure of climate conditions.

\begin{figure}[htb!]
	\centering
	\includegraphics[scale=0.28]{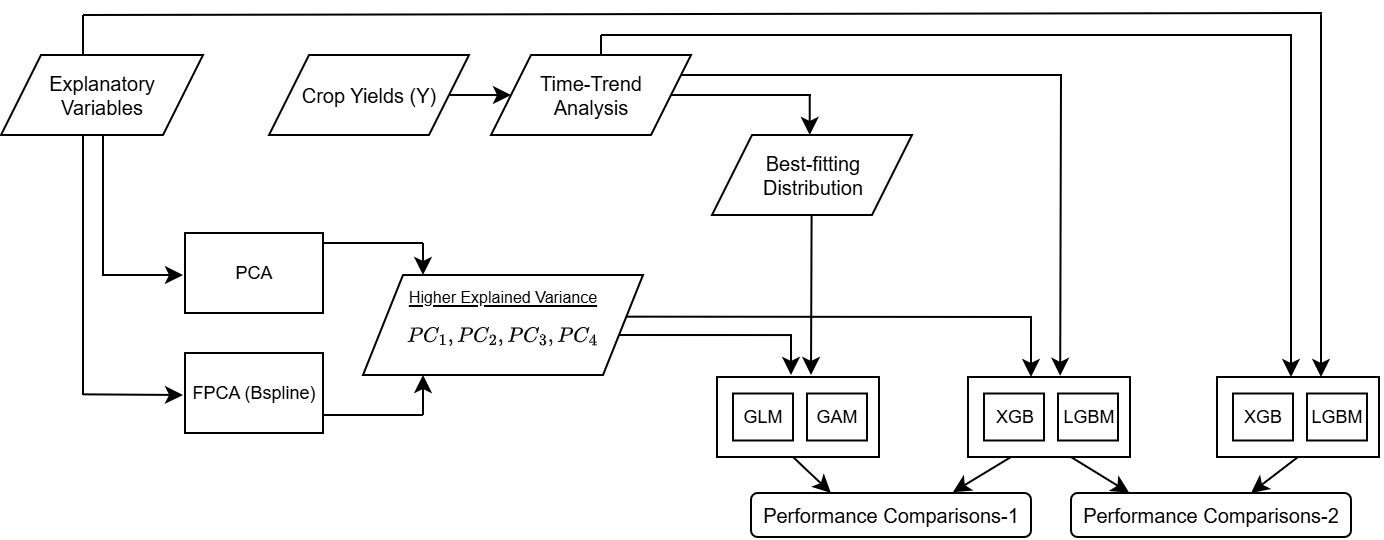}
	\caption{Flowchart of the proposed methodology}
	\label{fig1}
\end{figure}

A flowchart of the proposed methodology is presented in Figure \ref{fig1}. The framework consists of two distinct comparison structures designed to evaluate model performance under different input variables. In the first comparison analysis, the performance of regression models and machine learning algorithms is assessed using principal components derived from the explanatory variables. Since all indexes incorporated into the models are weather-based, they tend to reflect similar meteorological patterns, which may result in multicollinearity issues in regression analyses. To address this and simplify the explanatory data sets without compromising model accuracy, the PCA \citep{r5, r6} and FPCA \citep{r7, r8} are used as dimensionality reduction techniques.

Unlike traditional regression models, some machine learning algorithms are generally more robust to multicollinearity and do not strictly require prior data simplification for estimation. Nevertheless, feature selection or dimensionality reduction techniques can still be beneficial in enhancing model stability. In this context, the second comparison analysis incorporates the raw climate indexes directly into the machine learning algorithms without applying dimensionality reduction. This approach allows for a direct assessment of how machine learning methods handle potentially correlated climate variables, while also providing a basis for comparison with the principal component-based approach proposed in the methodology.

The Python packages \texttt{scikit-learn} \citep{r20} and \texttt{scikit-fda} \citep{r21} are used to implement PCA and FPCA with a B-spline basis, respectively. To maintain consistency and simplicity in model execution, the default settings of both packages are applied. From the transformed data sets, the first four principal components ($PC_1$, $PC_2$, $PC_3$, $PC_4$), which collectively account the majority of the variance in the data, are selected as the input variables for analyses.

\subsubsection{Time-Trend Analysis}\label{3.1.1.}

In crop yield analysis, detrending time series is essential, as agricultural production often follows a long-term upward trend driven by technological innovations, and advances in farming practices. Without appropriately removing or controlling for the trend in time, predictive models especially regression-based approaches may overstate their performance by simply capturing the persistent growth pattern instead of isolating the actual effects of weather and climate factors on yields.

A wide range of detrending techniques is available in the literature, including both parametric and non-parametric methods. Common approaches include linear trend \citep{r61}, quadratic trend \citep{r62}, moving average smoothing \citep{r63}, locally weighted regression \citep{r63}, and two-knot linear spline \citep{r60}. In this study, the linear detrending method, the most widely used approach in agricultural economics, is adopted, as it offers both simplicity and interpretability. The detrended crop yields ($Dy_t$) are defined as:

\begin{equation}\label{e91}
	Dy_t = y_t - \hat{y}_t + \hat{y}_0,
\end{equation}
where $y_t$ represents the observed crop yields, $\hat{y}_t$ denotes the values predicted from the fitted linear regression model, in which time $t$ is used as the independent variable (representing the temporal dimension of the data, e.g., monthly, seasonal, or annual observations). The addition of $\hat{y}_0$ ensures that all detrended yields are shifted to the level of the first predicted yield \citep{r1}. This adjustment prevents the occurrence of negative values in the detrended series, which is particularly important because GLM and GAM with some specific link functions cannot accommodate negative-valued inputs. From Equation (\ref{e91}), the removed time trends ($TT_t$) from the observed crop yields are obtained as follows:

\begin{equation}
	TT_t = y_t - Dy_t = \hat{y}_t - \hat{y}_0.
\end{equation}

Ultimately, the observed crop yield is expressed as the sum of two components: the detrended yield $Dy_t$, which captures the stationary variation driven by climate factors, and the time trend $TT_t$, which represents the systematic growth pattern over time. This decomposition ensures that the predictive models are applied to the detrended series, thereby isolating the effect of weather and climate variables on crop yields. The time stationarity of crop yields is evaluated using the Augmented Dickey-Fuller (ADF) test \citep{r64}, where the null hypothesis indicates that the crop yield series is non-stationary.

\subsubsection{Model Specification}\label{3.1.2.}

This subsection presents the modeling framework used in the study, which consists of both traditional statistical approaches and machine learning methods.

\paragraph{Regression-Based Models:}

After detrending the crop yield data, the next crucial step is to determine both the appropriate distribution of the yields and the most suitable link function for GLMs and GAMs before performing regression analyses. The Python package \texttt{distfit} \citep{r22} is used to identify the best-fitting distribution. This package evaluates 89 univariate distributions and selects the optimal one based on the residual sum of squares. For the link function, consistency across GLMs and GAMs is ensured by applying the log-link function in all models. The log-link function is particularly well-suited for crop yield data, as these values are inherently positive and continuous. The implementation of GLMs and GAMs is carried out using the R packages \texttt{stats} \citep{r53} and \texttt{mgcv} \citep{r54}, respectively. Following the detrending procedure, the detrended crop yields $Dy$ are used as the dependent variable in the subsequent GLM and GAM specifications. The GLM and GAM are expressed as:

\begin{equation}\label{e14}
	log(E(Dy|PC_1, PC_2, PC_3, PC_4)) = \beta_0 + \sum_{h=1}^{4} \beta_h * PC_h,
\end{equation}

\begin{equation}\label{e15}
	log(E(Dy|PC_1, PC_2, PC_3, PC_4)) = \beta_0 + \sum_{h=1}^{4} f_h(PC_h),
\end{equation}
respectively. Here, $PC_h$ $(h=1,2,3,4)$ denote the principal components. The coefficients $\beta_h$ are the model parameters, and $f_h(.)$ are smooth functions estimated using spline-based methods. In this study, penalized B-splines are used, as implemented in the \texttt{mgcv} package, with smoothing parameters automatically selected based on generalized cross-validation. In the analyses, both $Dy$ and $PC_h$ are treated as column vectors of size $N \times 1$, where $N$ corresponds to the number of observations (years).

\paragraph{Machine Learning Models:}

XGB and LGBM are both gradient boosting-based machine learning algorithms that build ensembles of decision trees in a sequential manner, where each new tree aims to improve the prediction errors of the previous ones. XGB incorporates regularization to control model complexity and reduce overfitting, while LGBM improves computational efficiency through a leaf-wise tree growth strategy and data sampling techniques. Both methods are well-suited for capturing nonlinear relationships and handling high-dimensional data.

To construct the XGB and LGBM models, the Python packages \texttt{xgboost} \citep{r55} and \texttt{lightgbm} \citep{r56} are chosen among the alternatives, respectively. To maintain consistency, the same hyper-parameters are used in all models. These are the number of estimators (500), maximum depth (2), and learning rate (0.01).

\subsubsection{Time Series Cross-Validation}\label{3.1.3.}

\begin{figure}[htb!]
	\centering
	\includegraphics[scale=0.65]{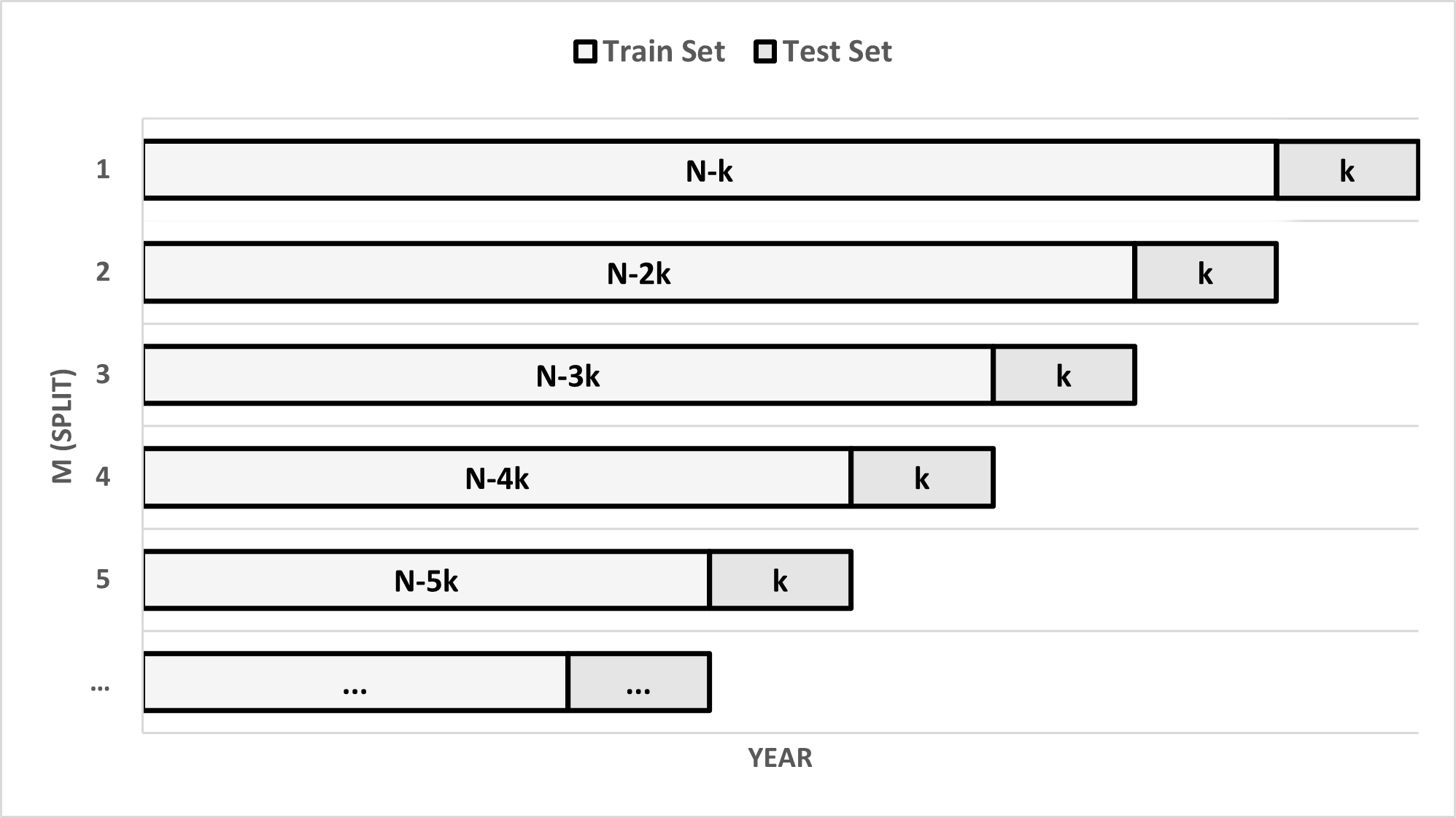}
	\caption{$M$-split leave-$k$-out cross-validation method}
	\label{fig1.1}
\end{figure}

Given the time series nature of climate indexes, training and test sets cannot be selected randomly when applying cross-validation methods. A random split would create the unrealistic situation in which future data points are used to predict past values. Therefore, we apply the iterative \textit{$M$-split leave-$k$-out cross-validation} method, which is specifically suitable for time series applications. In this approach, the data set is divided into $M$ sequential splits, where each test set contains a unique block of $k$ consecutive observations. By iteratively leaving out these $k$ observations for testing, while the preceding observations are used for training, the method ensures that the natural time order is preserved. The cross-validation procedure applied in this study is illustrated in Figure \ref{fig1.1}.

In the procedure, $M$ is the number of splits, $N$ denotes the total number of observations (years), while $k$ is set to approximately 10\% of $N$. Accordingly, the train/test share for the first iteration (split) is arranged as 90\% training data and 10\% testing data. For each split, the time stationarity of crop yields in the training data is first examined. If non-stationarity is detected, the yields are detrended using Equation (\ref{e91}), and both regression-based and machine-learning models are fitted to the resulting detrended training data. For performance evaluation, the removed time trends are reintroduced to the fitted values. In the case of the test set, the trend is extrapolated forward using only the estimates obtained from the training period, ensuring that no information from future observations is used in either model estimation or evaluation. Model performance is then evaluated on both training and test sets using the Mean Absolute Percentage Error (MAPE). The overall MAPE of the model across all splits for the training and test sets are:

\begin{equation}\label{e141}
	MAPE_{Train} = \frac{1}{M} \sum_{m=1}^{M} \bigg[ \frac{1}{N-mk} \sum_{t=1}^{N-mk} \frac{|y_t - (\hat{Dy}_t + TT_t)|}{y_t} \bigg],
\end{equation}

\begin{equation}\label{e151}
	MAPE_{Test} = \frac{1}{M} \sum_{m=1}^{M} \bigg[ \frac{1}{k} \sum_{t=N-mk+1}^{N-mk+k} \frac{|y_t - (\hat{Dy}_t + \hat{TT}^{(e)}_t)|}{y_t} \bigg],
\end{equation}
respectively. Here, $y_t$ denotes the observed crop yields, $\hat{Dy}_t$ represents the predicted detrended crop yields, $TT_t$ corresponds to the removed time trends, and $\hat{TT}^{(e)}_t$ shows extrapolated time trends, where $t$ indicates time (year). 

\subsection{Weather-Derivative Pricing}\label{3.2.}

For the financial market application, the similarities between WBIs and the components of the ACI are examined by evaluating the strength of their correlations and comparing their empirical cumulative distribution function (ECDF) values.

The comparison is structured across three groups (see Table \ref{tab1}). The time periods assigned to each group are carefully selected to align with the practical applications of climate indexes. Specifically, the summer period is used for warm-weather indexes (CDD-T90), reflecting their role in modeling cooling-related energy demand, while the winter period is considered for cold-weather indexes (HDD-T10), which are commonly used to assess heating requirements \citep{r15}. For precipitation-related indexes (PRE-P), the spring months are selected to capture the rainy season preceding the general harvest period of most crops, as precipitation timing plays a critical role in agricultural productivity \citep{r16, r17}. The dependence between the indexes is analyzed using the Spearman rank correlation coefficient \citep{r18}, while distributional comparisons are conducted through the Anderson-Darling (A-D) test \citep{r65}.

Mainly, the focus is on testing the performances of indexes within the weather-derivative framework and comparing their outcomes. For this purpose, the payoffs of weather derivative contracts are analyzed using underlying variables derived from both the WBIs and ACI components. This approach permits to investigate whether the indexes exhibit similar financial performance in market-based applications, thereby extending their potential utility beyond the agricultural sector.

\subsubsection{Derivative Contracts}\label{3.2.1.}

Weather derivatives are structured in several contract forms based on different underlying weather indexes. In the literature, the most commonly used indexes are CDD, HDD, and PRE \citep{r67, r2}. Regarding contract types, the most frequent forms are call/put options, and swaps \citep{r1, r68}. Basically, a call option provides a payout when the underlying weather index exceeds a predefined threshold, whereas a put option triggers a payout if the index falls below that threshold. Swaps, on the other hand, are structured as agreements between two parties to exchange cash flows based on the difference between the actual observed weather outcomes and an agreed-upon strike level. In this study, an uncapped call option is used, whose payoff $PO$ is defined as follows:

\begin{equation}\label{po}
	PO = \alpha \,  max\{I - K, 0\},
\end{equation}
where $I$ represents the value of the underlying weather index, and $K$ denotes the predetermined strike level. Here, $\alpha$ is the tick size, specifying the monetary value assigned per unit of the index.

\subsubsection{Pricing Methods}\label{3.2.2.}

Pricing of weather derivatives involves determining a fair contract value using one of several methods. Two commonly used ones, which are also adopted in this study, are the \textbf{Historical Burn Analysis} (HBA) and \textbf{Index Modeling} (IM) \citep{r70}. The HBA is widely used as a benchmark because of its simplicity, since it does not require fitting a probability distribution to the weather index or solving complex stochastic equations. The idea behind HBA is retrospective: it evaluates how the contract would have been performed in previous years and estimates the expected future payoff as the average of historical payoffs. The fair price of a call option under HBA can be computed with the following steps:

\begin{enumerate}\itemsep0em
	\item Determine the fair strike level ($K$) using historical weather index values.
	\item Calculate the historical payoffs of the option for each year based on the observed index values.
	\item Compute the mean of the historical payoffs, which serves as a fair price.
\end{enumerate}

Alternatively, within the IM method, the underlying weather index can be directly modeled to estimate the fair price of derivative contracts. The IM relies on statistical and stochastic representations of the weather index to construct its probability distribution, from which the expected payoff is derived. Once the distribution of the weather index is identified and its parameters are estimated, simulated index values are generated to compute expected payoffs. Similar to the HBA method, the mean of these simulated payoffs represents the fair price of the contract.

\section{Empirical Analysis} \label{sec4}

This section presents the empirical analysis conducted to evaluate the performance and applicability of the climate indexes across both agricultural and financial contexts. The primary objective is to assess the explanatory and predictive capabilities of the ACI in comparison with traditional WBIs, and to determine whether these indexes provide consistent insights across different application areas.

The section is organized as follows. Section \ref{4.1.} introduces the data sets used in the analysis, detailing their sources, structure, and preprocessing procedures. Section \ref{4.2.} addresses the agricultural application, where the predictive performance of the indexes is evaluated in the context of crop yield modeling. Section \ref{4.3.} presents the financial market application, examining the use of these indexes in weather derivative pricing and comparing their performance in terms of resulting payoffs and prices.

\clearpage

\subsection{Data} \label{4.1.}

The Actuaries Climate Index\textsuperscript{TM} data consist of time series for each component, either on a monthly basis or aggregated into meteorological seasons (three-month periods ending in February, May, August, and November). These time series, starting in 1961, are derived from measurements collected at meteorological stations across the U.S. and Canada. The data are available at different aggregation levels, including country-level and sub-regional series. This study uses the U.S. data, where states are grouped into seven sub-regions based on their geographic locations: Alaska (ALA), Central East Atlantic (CEA), Central West Pacific (CWP), Midwest (MID), Southeast Atlantic (SEA), Southern Plains (SPL), and Southwest Pacific (SWP). The analyses use standardized seasonal values for all six ACI components (T90, T10, P, D, W, S) as well as the composite index (ACI). This seasonal data, obtained from the official Actuaries Climate Index\textsuperscript{TM} website\footnote{ACI website: https://actuariesclimateindex.org/data/}, cover a 64-year period from 1961 to 2024 for all seven sub-regions.

Weather-based index (WBI) data are obtained from the National Oceanic and Atmospheric Administration (NOAA) Climate database\footnote{NOAA database: https://www.ncei.noaa.gov/access/monitoring/products/}, which provides monthly observations of CDD, HDD, and PRE for all U.S. states. To ensure consistency with the ACI data, these monthly series are aggregated into seasonal values for the period 1961-2024 by summing observations within each meteorological season. Sub-regional series are subsequently constructed by averaging state-level seasonal values within each region. However, due to the unavailability of CDD and HDD data for Alaska, the analyses are performed on the six remaining regions of the U.S.

Crop yield data are obtained from USDA’s National Agricultural Statistics Service (NASS) Quick Stats portal\footnote{NASS Quick Stats: https://quickstats.nass.usda.gov/}. Annual yields (in bushels per acre) for three major crops-corn (grain), wheat (winter), and soybeans-are collected at the state level for the period 1961-2024. Regional crop yields are constructed using acreage-weighted averages of state-level yields, where weights correspond to harvested acreage. The analyses for corn and wheat include all regions, whereas soybean data are limited to four regions (CEA, CWP, MID, and SEA) due to data availability constraints.

Overall, the empirical analysis is conducted using temporally aligned data sets over the period 1961-2024, where weather indexes are aggregated to the seasonal level and crop yields are observed annually. This corresponds to 64 annual observations per region for crop yields and four seasonal observations per year for each weather index series. These preprocessing steps ensure consistency across data sets and enable a coherent evaluation of climate impacts across agricultural and financial applications.

\subsection{Agricultural Case} \label{4.2.}

\begin{figure}[htb!]
	\centering
	\includegraphics[scale=0.5]{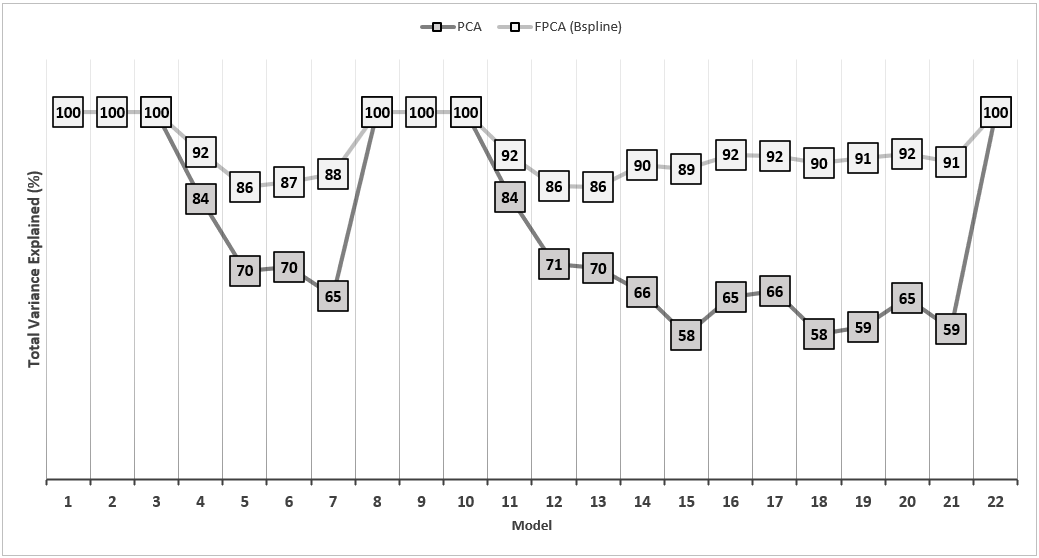}
	\caption{Explained variance comparisons of PCA and FPCA as regional averages}
	\label{fig2}
\end{figure}

Firstly, the explanatory variables in each model (Table \ref{tab2}) are transformed into their principal components using PCA and FPCA. Figure \ref{fig2} illustrates the total variation in the data sets, represented as regional averages, explained by the first four principal components derived from both methods for each model. The results demonstrate that the total variance explained by FPCA consistently surpasses that of PCA across all models, highlighting FPCA’s superior performance when dealing with time-dependent data sets like climate indexes.

Upon closer examination of Figure \ref{fig2}, the models (1, 2, 3, 8, 9, 10, and 22) show a total explained variance of 100\% for both PCA and FPCA. This is expected, as these models contain only one weather variable represented by its four seasonal values. Consequently, PCA and FPCA serve purely as data transformation tools for these cases without affecting the total variance explained.

\begin{figure}[hbt!]
	\begin{adjustwidth}{0cm}{0cm}
		\centering
	\includegraphics[width=8.0cm]{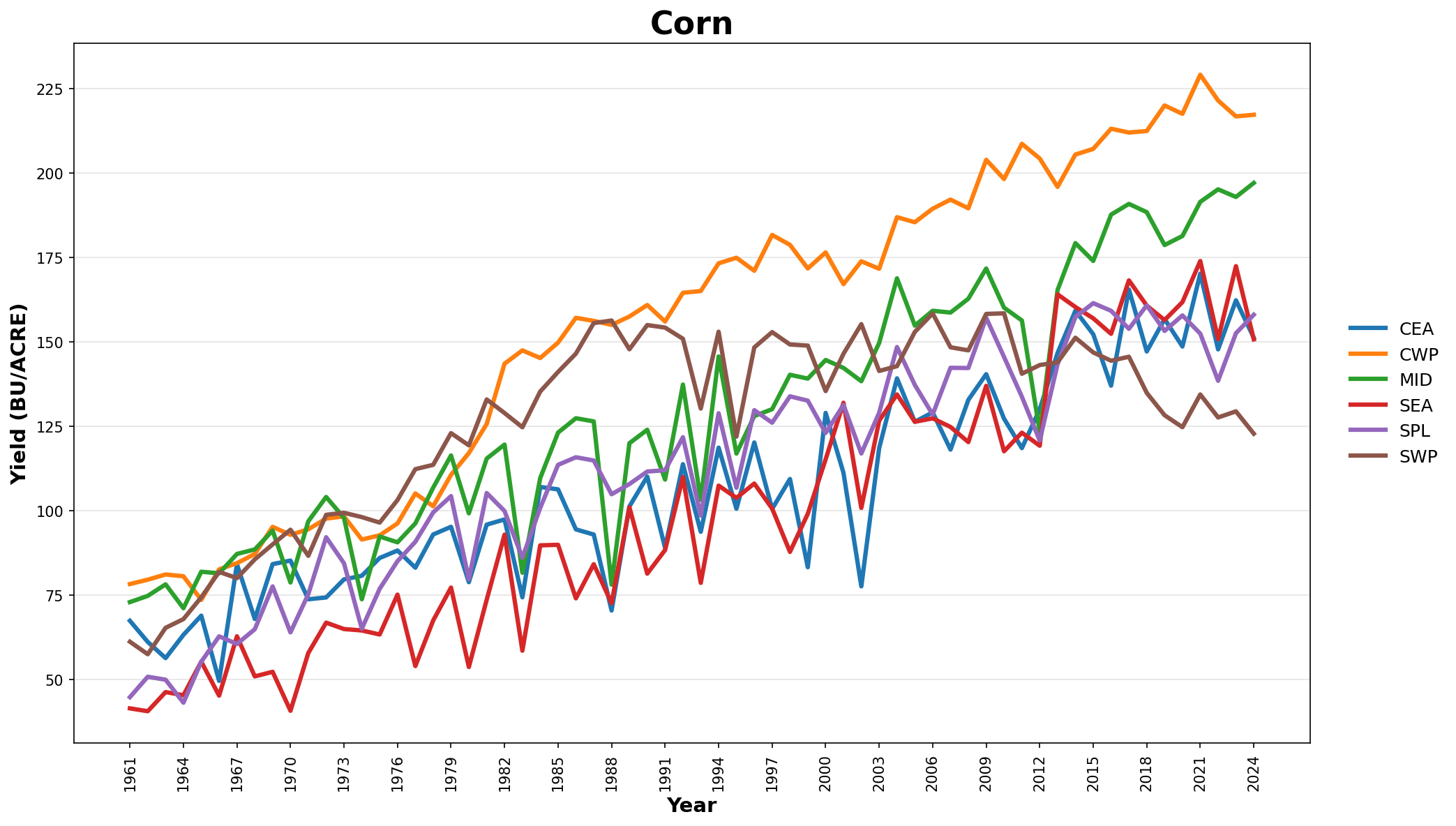}
	\includegraphics[width=8.0cm]{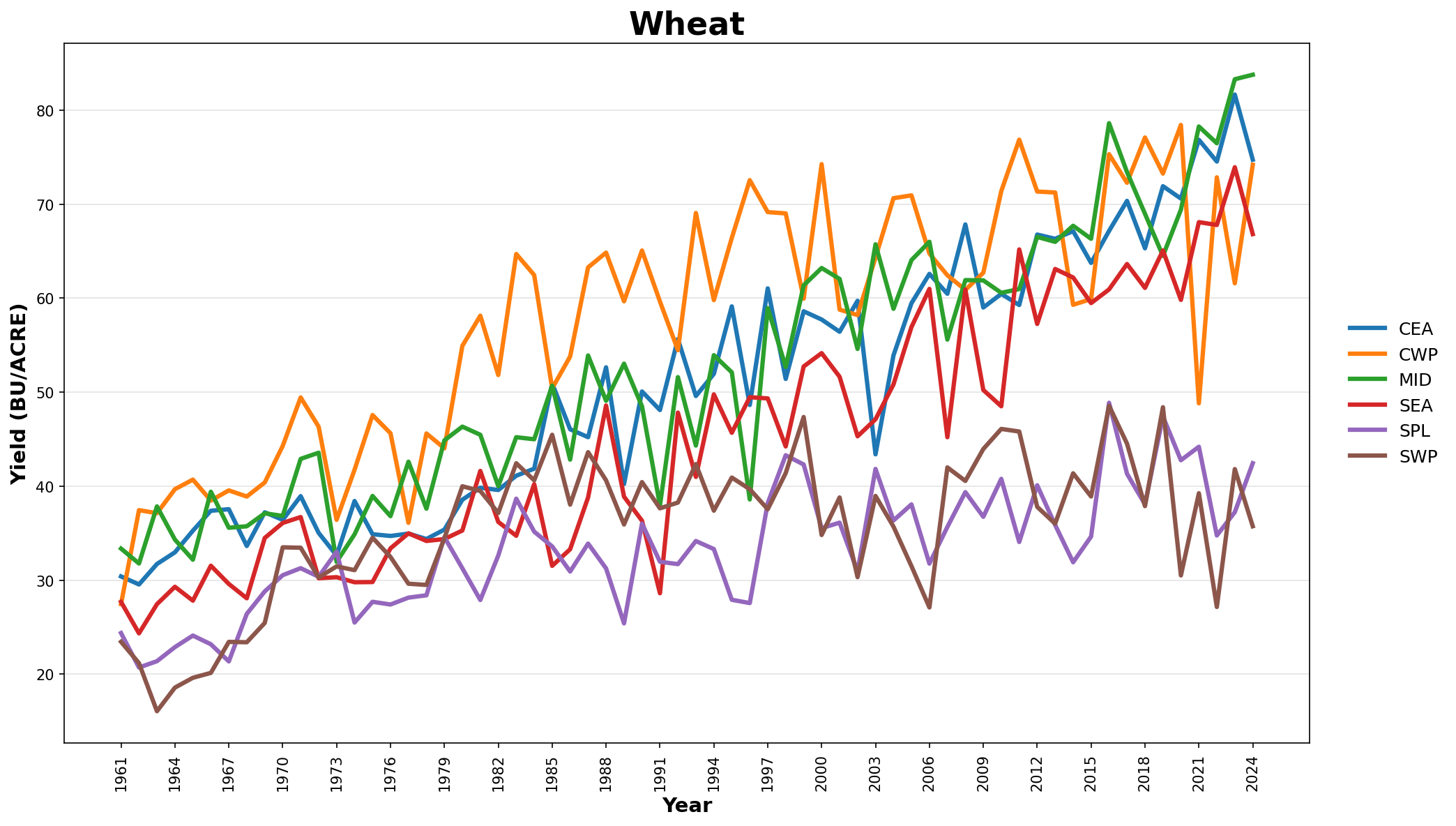}\\
	\includegraphics[width=8.0cm]{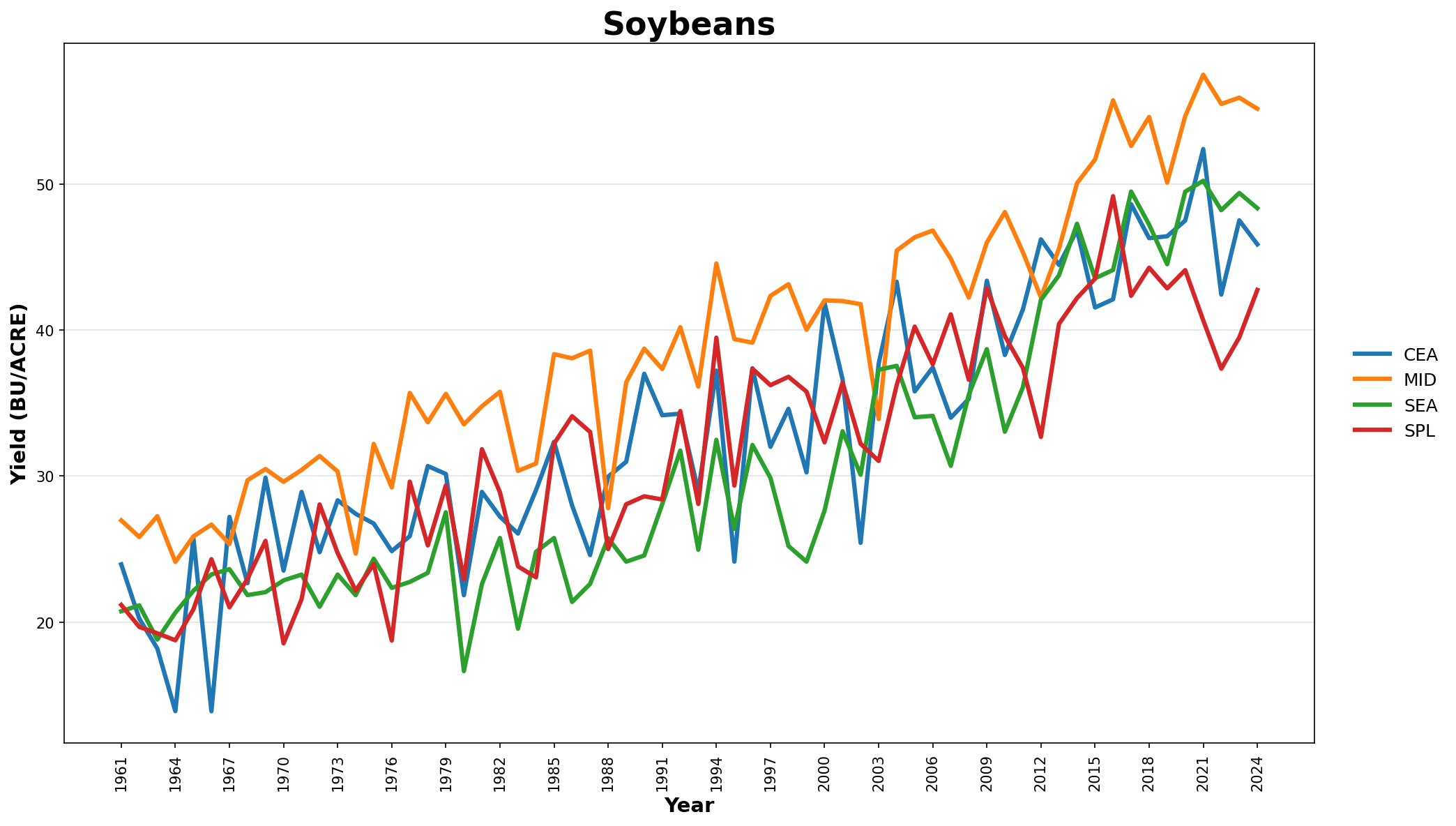}
\end{adjustwidth}
\caption{Crop yields over time}
\label{fig41}
\end{figure}

For a detailed comparison, recall that the first seven models are constructed based on WBIs, while the subsequent seven models (Models 8-14) use their ACI counterparts. When comparing the total explained variance between these paired models (e.g., 1 vs. 8, 2 vs. 9, 3 vs. 10, and so forth), it is clear that the values are nearly equivalent. For instance, PCA explains 70\% of the variance in Model 5, whereas it explains 71\% in Model 12. Similarly, FPCA captures 88\% of the variance in Model 7 and 90\% in Model 14. Since the choice of explanatory variables, and consequently the derived principal components, directly affects the outcomes of the analyses, ensuring high and comparable explained variance values across both index groups is crucial for robustness. Given that FPCA consistently accounts for a larger proportion of variance across all models, with values exceeding 85\%, the FPCA-derived principal components are selected for the subsequent analyses, ensuring better model accuracy and reliability. The crop yields used as dependent variables in the regression and machine learning analyses are illustrated for each region in Figure \ref{fig41}.

The GLM and GAM models used in the empirical analysis follow the specifications given in Equations (\ref{e14}) and (\ref{e15}), respectively. For each model, predictions are obtained based on the detrended crop yields as the dependent variables and the explanatory variables as inputs (see Table \ref{tab2} and Table \ref{tab22}). Model performance is then evaluated using the MAPE, computed according to Equations (\ref{e141}) and (\ref{e151}) for the training and test sets within the cross-validation framework. The parameters for cross-validation are set as the number of years $N=64$, the number of splits $M=5$, and the leave-$k$-out cross-validation parameter $k=6$ in Equations (\ref{e141}) and (\ref{e151}).

\begin{figure}[htb!]
	\begin{adjustwidth}{0.5cm}{0cm}
		\centering
		\includegraphics[width=13.5cm]{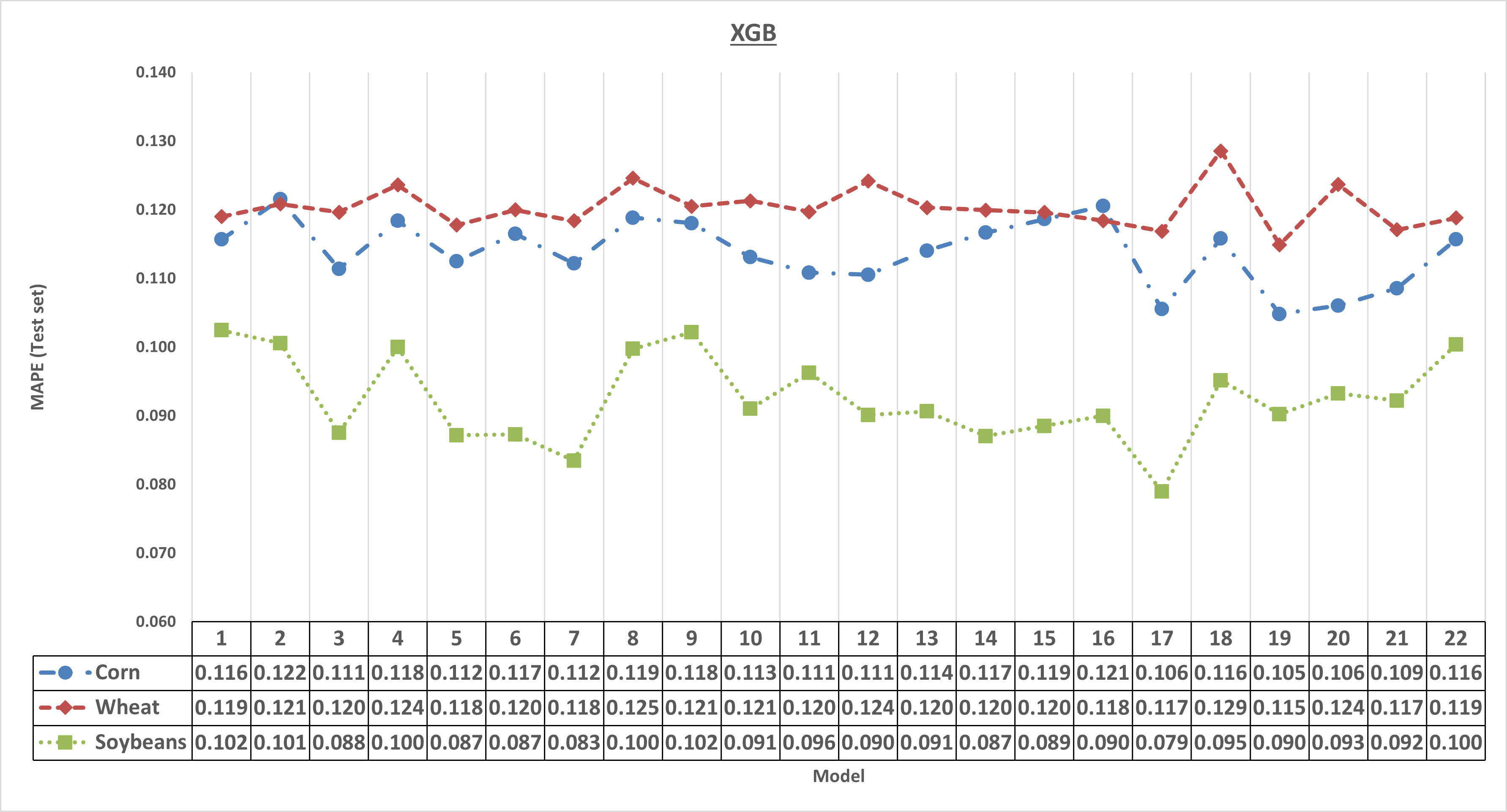}
	\end{adjustwidth}
	\caption{MAPE values as regional averages for each crop and model with XGB (Test set)}
	\label{fig5}
\end{figure}

Although MAPE is used as the primary performance metric, additional robustness checks using Mean Absolute Error and Root Mean Square Error yield qualitatively similar results and are therefore omitted for brevity. The MAPE values as regional averages for the training data of each crop are presented in Table \ref{tab51}. Based on the average values presented in the last row of Table \ref{tab51}, the GAM performs slightly better than the GLM, achieving lower MAPE values, though the difference remains relatively small. A similar pattern is observed between the XGB and LGBM, where XGB consistently outperforms LGBM, ultimately producing the lowest MAPE values among all methods. For the training data set, the average error rate of XGB across all crops is around 4\%, whereas the corresponding error rates for the other three methods range between 8\% and 11\%. Since XGB consistently outperforms GLM, GAM, and LGBM for the training set, the MAPE values as regional averages for the test set are reported only for XGB in Figure \ref{fig5}. However, for a more comprehensive comparison across methods, the regional MAPE values computed separately for both training and test data sets for all crops are given in Tables \ref{tab6}-\ref{tab82} of the Appendix.

\begin{table}[hbt!]
\centering
\caption{MAPE values as regional averages for each crop, model and method  (Train set)}
\label{tab51}
\begin{adjustwidth}{1.5cm}{0cm}
	\scalebox{0.77}{
		\begin{tabular}{|c|cccc|cccc|cccc|}
			\hline
			\multirow{2}{*}{\textbf{Model}} & \multicolumn{4}{c|}{\textbf{Corn}}                                                                                               & \multicolumn{4}{c|}{\textbf{Wheat}}                                                                                              & \multicolumn{4}{c|}{\textbf{Soybeans}}                                                                                           \\ \cline{2-13} 
			& \multicolumn{1}{c|}{\textbf{GLM}}   & \multicolumn{1}{c|}{\textbf{GAM}}   & \multicolumn{1}{c|}{\textbf{XGB}}   & \textbf{LGBM}  & \multicolumn{1}{c|}{\textbf{GLM}}   & \multicolumn{1}{c|}{\textbf{GAM}}   & \multicolumn{1}{c|}{\textbf{XGB}}   & \textbf{LGBM}  & \multicolumn{1}{c|}{\textbf{GLM}}   & \multicolumn{1}{c|}{\textbf{GAM}}   & \multicolumn{1}{c|}{\textbf{XGB}}   & \textbf{LGBM}  \\ \hline
			\textbf{1}                      & \multicolumn{1}{c|}{0.107}          & \multicolumn{1}{c|}{0.093}          & \multicolumn{1}{c|}{0.047}          & 0.094          & \multicolumn{1}{c|}{0.095}          & \multicolumn{1}{c|}{0.080}          & \multicolumn{1}{c|}{0.052}          & 0.094          & \multicolumn{1}{c|}{0.119}          & \multicolumn{1}{c|}{0.103}          & \multicolumn{1}{c|}{0.043}          & 0.087          \\ \hline
			\textbf{2}                      & \multicolumn{1}{c|}{0.106}          & \multicolumn{1}{c|}{0.095}          & \multicolumn{1}{c|}{0.051}          & 0.096          & \multicolumn{1}{c|}{0.096}          & \multicolumn{1}{c|}{0.081}          & \multicolumn{1}{c|}{0.051}          & 0.094          & \multicolumn{1}{c|}{0.121}          & \multicolumn{1}{c|}{0.115}          & \multicolumn{1}{c|}{0.048}          & 0.090          \\ \hline
			\textbf{3}                      & \multicolumn{1}{c|}{0.102}          & \multicolumn{1}{c|}{0.084}          & \multicolumn{1}{c|}{0.048}          & 0.094          & \multicolumn{1}{c|}{0.091}          & \multicolumn{1}{c|}{0.081}          & \multicolumn{1}{c|}{0.048}          & 0.089          & \multicolumn{1}{c|}{0.114}          & \multicolumn{1}{c|}{0.104}          & \multicolumn{1}{c|}{0.037}          & 0.082          \\ \hline
			\textbf{4}                      & \multicolumn{1}{c|}{0.105}          & \multicolumn{1}{c|}{0.092}          & \multicolumn{1}{c|}{0.047}          & 0.093          & \multicolumn{1}{c|}{0.095}          & \multicolumn{1}{c|}{0.084}          & \multicolumn{1}{c|}{0.052}          & 0.094          & \multicolumn{1}{c|}{0.116}          & \multicolumn{1}{c|}{0.111}          & \multicolumn{1}{c|}{0.047}          & 0.088          \\ \hline
			\textbf{5}                      & \multicolumn{1}{c|}{0.101}          & \multicolumn{1}{c|}{0.083}          & \multicolumn{1}{c|}{0.046}          & 0.093          & \multicolumn{1}{c|}{0.090}          & \multicolumn{1}{c|}{0.084}          & \multicolumn{1}{c|}{0.049}          & 0.090          & \multicolumn{1}{c|}{0.110}          & \multicolumn{1}{c|}{0.097}          & \multicolumn{1}{c|}{0.037}          & 0.082          \\ \hline
			\textbf{6}                      & \multicolumn{1}{c|}{0.102}          & \multicolumn{1}{c|}{0.084}          & \multicolumn{1}{c|}{0.045}          & 0.093          & \multicolumn{1}{c|}{0.089}          & \multicolumn{1}{c|}{0.079}          & \multicolumn{1}{c|}{0.047}          & 0.089          & \multicolumn{1}{c|}{0.109}          & \multicolumn{1}{c|}{0.081}          & \multicolumn{1}{c|}{0.038}          & 0.085          \\ \hline
			\textbf{7}                      & \multicolumn{1}{c|}{0.102}          & \multicolumn{1}{c|}{0.084}          & \multicolumn{1}{c|}{0.047}          & 0.093          & \multicolumn{1}{c|}{0.088}          & \multicolumn{1}{c|}{0.075}          & \multicolumn{1}{c|}{0.046}          & 0.089          & \multicolumn{1}{c|}{0.109}          & \multicolumn{1}{c|}{0.087}          & \multicolumn{1}{c|}{0.039}          & 0.084          \\ \hline
			\textbf{8}                      & \multicolumn{1}{c|}{0.106}          & \multicolumn{1}{c|}{0.092}          & \multicolumn{1}{c|}{0.049}          & 0.093          & \multicolumn{1}{c|}{0.095}          & \multicolumn{1}{c|}{0.082}          & \multicolumn{1}{c|}{0.051}          & 0.093          & \multicolumn{1}{c|}{0.119}          & \multicolumn{1}{c|}{0.104}          & \multicolumn{1}{c|}{0.044}          & 0.085          \\ \hline
			\textbf{9}                      & \multicolumn{1}{c|}{0.107}          & \multicolumn{1}{c|}{0.094}          & \multicolumn{1}{c|}{0.048}          & 0.095          & \multicolumn{1}{c|}{0.099}          & \multicolumn{1}{c|}{0.090}          & \multicolumn{1}{c|}{0.052}          & 0.094          & \multicolumn{1}{c|}{0.124}          & \multicolumn{1}{c|}{0.102}          & \multicolumn{1}{c|}{0.049}          & 0.089          \\ \hline
			\textbf{10}                     & \multicolumn{1}{c|}{0.101}          & \multicolumn{1}{c|}{0.089}          & \multicolumn{1}{c|}{0.047}          & 0.093          & \multicolumn{1}{c|}{0.092}          & \multicolumn{1}{c|}{0.087}          & \multicolumn{1}{c|}{0.046}          & 0.089          & \multicolumn{1}{c|}{0.110}          & \multicolumn{1}{c|}{0.098}          & \multicolumn{1}{c|}{0.044}          & 0.087          \\ \hline
			\textbf{11}                     & \multicolumn{1}{c|}{0.105}          & \multicolumn{1}{c|}{0.089}          & \multicolumn{1}{c|}{0.048}          & 0.094          & \multicolumn{1}{c|}{0.096}          & \multicolumn{1}{c|}{0.081}          & \multicolumn{1}{c|}{0.050}          & 0.093          & \multicolumn{1}{c|}{0.121}          & \multicolumn{1}{c|}{0.102}          & \multicolumn{1}{c|}{0.048}          & 0.088          \\ \hline
			\textbf{12}                     & \multicolumn{1}{c|}{0.098}          & \multicolumn{1}{c|}{0.085}          & \multicolumn{1}{c|}{0.044}          & 0.090          & \multicolumn{1}{c|}{0.089}          & \multicolumn{1}{c|}{0.077}          & \multicolumn{1}{c|}{0.048}          & 0.090          & \multicolumn{1}{c|}{0.103}          & \multicolumn{1}{c|}{0.082}          & \multicolumn{1}{c|}{0.042}          & 0.084          \\ \hline
			\textbf{13}                     & \multicolumn{1}{c|}{0.101}          & \multicolumn{1}{c|}{0.089}          & \multicolumn{1}{c|}{0.047}          & 0.091          & \multicolumn{1}{c|}{0.092}          & \multicolumn{1}{c|}{0.082}          & \multicolumn{1}{c|}{0.049}          & 0.091          & \multicolumn{1}{c|}{0.110}          & \multicolumn{1}{c|}{0.098}          & \multicolumn{1}{c|}{0.045}          & 0.086          \\ \hline
			\textbf{14}                     & \multicolumn{1}{c|}{0.101}          & \multicolumn{1}{c|}{0.089}          & \multicolumn{1}{c|}{0.047}          & 0.092          & \multicolumn{1}{c|}{0.090}          & \multicolumn{1}{c|}{0.072}          & \multicolumn{1}{c|}{0.046}          & 0.090          & \multicolumn{1}{c|}{0.106}          & \multicolumn{1}{c|}{0.095}          & \multicolumn{1}{c|}{0.046}          & 0.087          \\ \hline
			\textbf{15}                     & \multicolumn{1}{c|}{0.099}          & \multicolumn{1}{c|}{0.090}          & \multicolumn{1}{c|}{0.047}          & 0.094          & \multicolumn{1}{c|}{0.092}          & \multicolumn{1}{c|}{0.078}          & \multicolumn{1}{c|}{0.049}          & 0.092          & \multicolumn{1}{c|}{0.107}          & \multicolumn{1}{c|}{0.087}          & \multicolumn{1}{c|}{0.043}          & 0.086          \\ \hline
			\textbf{16}                     & \multicolumn{1}{c|}{0.102}          & \multicolumn{1}{c|}{0.084}          & \multicolumn{1}{c|}{0.046}          & 0.092          & \multicolumn{1}{c|}{0.093}          & \multicolumn{1}{c|}{0.079}          & \multicolumn{1}{c|}{0.046}          & 0.089          & \multicolumn{1}{c|}{0.107}          & \multicolumn{1}{c|}{0.092}          & \multicolumn{1}{c|}{0.045}          & 0.089          \\ \hline
			\textbf{17}                     & \multicolumn{1}{c|}{0.094}          & \multicolumn{1}{c|}{0.083}          & \multicolumn{1}{c|}{0.045}          & 0.090          & \multicolumn{1}{c|}{0.085}          & \multicolumn{1}{c|}{0.073}          & \multicolumn{1}{c|}{0.044}          & 0.089          & \multicolumn{1}{c|}{0.100}          & \multicolumn{1}{c|}{0.089}          & \multicolumn{1}{c|}{0.047}          & 0.088          \\ \hline
			\textbf{18}                     & \multicolumn{1}{c|}{0.102}          & \multicolumn{1}{c|}{0.085}          & \multicolumn{1}{c|}{0.049}          & 0.094          & \multicolumn{1}{c|}{0.093}          & \multicolumn{1}{c|}{0.077}          & \multicolumn{1}{c|}{0.051}          & 0.090          & \multicolumn{1}{c|}{0.109}          & \multicolumn{1}{c|}{0.097}          & \multicolumn{1}{c|}{0.047}          & 0.088          \\ \hline
			\textbf{19}                     & \multicolumn{1}{c|}{0.093}          & \multicolumn{1}{c|}{0.080}          & \multicolumn{1}{c|}{0.042}          & 0.089          & \multicolumn{1}{c|}{0.087}          & \multicolumn{1}{c|}{0.077}          & \multicolumn{1}{c|}{0.046}          & 0.088          & \multicolumn{1}{c|}{0.097}          & \multicolumn{1}{c|}{0.084}          & \multicolumn{1}{c|}{0.046}          & 0.088          \\ \hline
			\textbf{20}                     & \multicolumn{1}{c|}{0.096}          & \multicolumn{1}{c|}{0.083}          & \multicolumn{1}{c|}{0.044}          & 0.092          & \multicolumn{1}{c|}{0.088}          & \multicolumn{1}{c|}{0.074}          & \multicolumn{1}{c|}{0.046}          & 0.089          & \multicolumn{1}{c|}{0.099}          & \multicolumn{1}{c|}{0.086}          & \multicolumn{1}{c|}{0.044}          & 0.088          \\ \hline
			\textbf{21}                     & \multicolumn{1}{c|}{0.096}          & \multicolumn{1}{c|}{0.077}          & \multicolumn{1}{c|}{0.043}          & 0.090          & \multicolumn{1}{c|}{0.088}          & \multicolumn{1}{c|}{0.080}          & \multicolumn{1}{c|}{0.050}          & 0.088          & \multicolumn{1}{c|}{0.101}          & \multicolumn{1}{c|}{0.091}          & \multicolumn{1}{c|}{0.049}          & 0.089          \\ \hline
			\textbf{22}                     & \multicolumn{1}{c|}{0.105}          & \multicolumn{1}{c|}{0.089}          & \multicolumn{1}{c|}{0.050}          & 0.095          & \multicolumn{1}{c|}{0.093}          & \multicolumn{1}{c|}{0.086}          & \multicolumn{1}{c|}{0.050}          & 0.092          & \multicolumn{1}{c|}{0.118}          & \multicolumn{1}{c|}{0.102}          & \multicolumn{1}{c|}{0.053}          & 0.091          \\ \hline
			\textbf{Avg.}                   & \multicolumn{1}{c|}{\textbf{0.102}} & \multicolumn{1}{c|}{\textbf{0.087}} & \multicolumn{1}{c|}{\textbf{0.047}} & \textbf{0.093} & \multicolumn{1}{c|}{\textbf{0.092}} & \multicolumn{1}{c|}{\textbf{0.080}} & \multicolumn{1}{c|}{\textbf{0.049}} & \textbf{0.091} & \multicolumn{1}{c|}{\textbf{0.110}} & \multicolumn{1}{c|}{\textbf{0.096}} & \multicolumn{1}{c|}{\textbf{0.045}} & \textbf{0.087} \\ \hline
	\end{tabular}}
\end{adjustwidth}
\end{table}

In Figure \ref{fig5}, Model 19 achieves the best performance with a MAPE of 10.5\% for corn. The results of each model facilitate a detailed examination of how individual ACI components influence crop yields. For instance, Model 14, which includes temperature and precipitation components (T90, T10, P), achieves a MAPE of 11.7\%. Sequentially adding the wind (W), drought (D), and sea-level (S) components in Models 15, 16, and 17, respectively, allows the evaluation of their incremental effects. Among these, Model 17 produces the best result with a MAPE of 10.6\%, indicating that the sea-level component has the strongest positive effect on corn yield prediction, reducing the error from 11.7\% to 10.6\%. Furthermore, when the wind and sea-level components are combined (Model 19), model performance improves slightly further, reaching 10.5\% MAPE. Interestingly, Model 21, which incorporates all six ACI components, performs worse than Model 19. This outcome suggests that while wind and sea-level components enhance predictive accuracy, the inclusion of the drought has a diminishing effect on model performance when all other five climate variables of ACI are already used as predictors.

For wheat, similar to the findings for corn, Model 19 achieves the best overall performance, yielding a MAPE of 11.5\%. When examining the individual component models, Model 8 (T90), Model 9 (T10), and Model 10 (P) produce MAPE values of 12.5\%, 12.1\%, and 12.1\%, respectively. These results indicate that, when analyzed individually, precipitation has a relatively higher impact on wheat yield prediction than the temperature-related components. When precipitation is combined with low-temperature extremes (as in Model 13), model’s performance improves slightly, with the MAPE decreasing to 12.0\%. Furthermore, considering Models 15 (12.0\%), 16 (11.8\%), and 17 (11.7\%), which sequentially incorporate wind, drought, and sea-level components, alongside temperature and precipitation, the results reveal that the sea-level component contributes the most substantial improvement in predictive accuracy.

For soybeans, the results differ slightly from those obtained for corn and wheat. The best-performing model for soybeans is Model 17 (MAPE of 7.9\%). Unlike the findings for the other two crops, the inclusion of the wind component appears to negatively affect model performance for soybean yield. The performance difference between Model 12 (T90 and P) and Model 13 (T10 and P) is noteworthy. Model 12 achieves a MAPE of 9.0\%, while Model 13 yields a higher error of 9.1\%, indicating that warm weather extremes (T90) appear to have a stronger influence on soybean yield than cold weather extremes (T10). This conclusion is further supported by the performance results of Model 8 (T90: 10.0\%) and Model 9 (T10: 10.2\%).

When analyzing the best-performing models for the three crops (Model 19 for corn and wheat, Model 17 for soybeans), it becomes clear that, alongside the temperature and precipitation components, sea-level change plays a significant role in yield prediction. This finding is also consistent with the agronomic and climate-impact literature, which document that sea-level rise can affect agricultural productivity more broadly though mechanisms such as increased soil salinity, flooding, and the loss of agricultural land \citep{r24, r72, r73}. Furthermore, wind appears to have a less pronounced effect on soybean yield compared to corn and wheat, likely due to the relatively short stature of soybean plants, which makes them more resistant to wind-induced damage. In contrast, taller crops like corn and wheat are more vulnerable to lodging\footnote{Lodging refers to the bending or collapse of crop stems, often caused by strong winds, which makes harvesting difficult.} and other wind-related stresses that can negatively impact their overall yield.

A detailed comparison of the performance results for paired models (1 vs. 8, 2 vs. 9, 3 vs. 10, and so forth) on crop yields also reveals consistent and satisfactory outcomes. For instance, the performance difference between Model 1 (CDD: 11.6\%) and Model 8 (T90: 11.9\%) for corn, Model 2 (HDD: 12.1\%) and Model 9 (T10: 12.1\%) for wheat, Model 3 (PRE: 8.8\%) and Model 10 (P: 9.1\%) for soybeans are almost negligible, suggesting comparable explanatory power between ACI components and their WBI counterparts.

Another critical observation is that Model 22, which uses the composite ACI by aggregating all six components into a single index, consistently underperforms compared to Model 21, which incorporates each component separately, across all crops. These results emphasize that the composite ACI alone fails to fully capture and explain the complex weather conditions affecting crop yields. Consequently, analyzing the individual components separately provides a more nuanced and effective modeling approach, leading to improved accuracy in predicting yield variations.

\begin{table}[hbt!]
\centering
\caption{Performance Comparison-2: MAPE values as regional averages for each crop (Test set)}
\label{tab11}
\scalebox{0.79}{
	\begin{tabular}{|c|cc|cc|cc|}
		\hline
		\multirow{2}{*}{\textbf{Model}} & \multicolumn{2}{c|}{\textbf{Corn}}                 & \multicolumn{2}{c|}{\textbf{Wheat}}                & \multicolumn{2}{c|}{\textbf{Soybeans}}             \\ \cline{2-7} 
		& \multicolumn{1}{c|}{\textbf{XGB}} & \textbf{XGB-C} & \multicolumn{1}{c|}{\textbf{XGB}} & \textbf{XGB-C} & \multicolumn{1}{c|}{\textbf{XGB}} & \textbf{XGB-C} \\ \hline
		\textbf{1}                      & \multicolumn{1}{c|}{0.116}        & 0.115          & \multicolumn{1}{c|}{0.119}        & 0.117          & \multicolumn{1}{c|}{0.102}        & 0.099          \\ \hline
		\textbf{2}                      & \multicolumn{1}{c|}{0.122}        & 0.129          & \multicolumn{1}{c|}{0.121}        & 0.118          & \multicolumn{1}{c|}{0.101}        & 0.103          \\ \hline
		\textbf{3}                      & \multicolumn{1}{c|}{0.111}        & 0.113          & \multicolumn{1}{c|}{0.120}        & 0.124          & \multicolumn{1}{c|}{0.088}        & 0.083          \\ \hline
		\textbf{4}                      & \multicolumn{1}{c|}{0.118}        & 0.115          & \multicolumn{1}{c|}{0.124}        & 0.112          & \multicolumn{1}{c|}{0.100}        & 0.099          \\ \hline
		\textbf{5}                      & \multicolumn{1}{c|}{0.112}        & 0.110          & \multicolumn{1}{c|}{0.118}        & 0.111          & \multicolumn{1}{c|}{0.087}        & 0.080          \\ \hline
		\textbf{6}                      & \multicolumn{1}{c|}{0.117}        & 0.117          & \multicolumn{1}{c|}{0.120}        & 0.111          & \multicolumn{1}{c|}{0.087}        & 0.085          \\ \hline
		\textbf{7}                      & \multicolumn{1}{c|}{0.112}        & 0.112          & \multicolumn{1}{c|}{0.118}        & 0.111          & \multicolumn{1}{c|}{0.083}        & 0.080          \\ \hline
		\textbf{8}                      & \multicolumn{1}{c|}{0.119}        & 0.116          & \multicolumn{1}{c|}{0.125}        & 0.125          & \multicolumn{1}{c|}{0.100}        & 0.100          \\ \hline
		\textbf{9}                      & \multicolumn{1}{c|}{0.118}        & 0.125          & \multicolumn{1}{c|}{0.121}        & 0.119          & \multicolumn{1}{c|}{0.102}        & 0.104          \\ \hline
		\textbf{10}                     & \multicolumn{1}{c|}{0.113}        & 0.112          & \multicolumn{1}{c|}{0.121}        & 0.115          & \multicolumn{1}{c|}{0.091}        & 0.088          \\ \hline
		\textbf{11}                     & \multicolumn{1}{c|}{0.111}        & 0.118          & \multicolumn{1}{c|}{0.120}        & 0.112          & \multicolumn{1}{c|}{0.096}        & 0.100          \\ \hline
		\textbf{12}                     & \multicolumn{1}{c|}{0.111}        & 0.108          & \multicolumn{1}{c|}{0.124}        & 0.113          & \multicolumn{1}{c|}{0.090}        & 0.090          \\ \hline
		\textbf{13}                     & \multicolumn{1}{c|}{0.114}        & 0.111          & \multicolumn{1}{c|}{0.120}        & 0.118          & \multicolumn{1}{c|}{0.091}        & 0.091          \\ \hline
		\textbf{14}                     & \multicolumn{1}{c|}{0.117}        & 0.112          & \multicolumn{1}{c|}{0.120}        & 0.114          & \multicolumn{1}{c|}{0.087}        & 0.092          \\ \hline
		\textbf{15}                     & \multicolumn{1}{c|}{0.119}        & 0.111          & \multicolumn{1}{c|}{0.120}        & 0.115          & \multicolumn{1}{c|}{0.089}        & 0.090          \\ \hline
		\textbf{16}                     & \multicolumn{1}{c|}{0.121}        & 0.114          & \multicolumn{1}{c|}{0.118}        & 0.113          & \multicolumn{1}{c|}{0.090}        & 0.092          \\ \hline
		\textbf{17}                     & \multicolumn{1}{c|}{0.106}        & 0.106          & \multicolumn{1}{c|}{0.117}        & 0.107          & \multicolumn{1}{c|}{\textbf{0.079$^*$}}        & \textbf{0.079$^\phi$}          \\ \hline
		\textbf{18}                     & \multicolumn{1}{c|}{0.116}        & 0.112          & \multicolumn{1}{c|}{0.129}        & 0.112          & \multicolumn{1}{c|}{0.095}        & 0.089          \\ \hline
		\textbf{19}                     & \multicolumn{1}{c|}{\textbf{0.105$^*$}}        & \textbf{0.101$^\phi$}          & \multicolumn{1}{c|}{\textbf{0.115$^*$}}        & \textbf{0.105$^\phi$}          & \multicolumn{1}{c|}{0.090}        & 0.083          \\ \hline
		\textbf{20}                     & \multicolumn{1}{c|}{0.106}        & 0.104          & \multicolumn{1}{c|}{0.124}        & 0.109          & \multicolumn{1}{c|}{0.093}        & 0.083          \\ \hline
		\textbf{21}                     & \multicolumn{1}{c|}{0.109}        & \textbf{0.100$^*$}          & \multicolumn{1}{c|}{0.117}        & \textbf{0.104$^*$}          & \multicolumn{1}{c|}{0.092}        & \textbf{0.075$^*$}          \\ \hline
		\textbf{22}                     & \multicolumn{1}{c|}{0.116}        & 0.110          & \multicolumn{1}{c|}{0.119}        & 0.115          & \multicolumn{1}{c|}{0.100}        & 0.104          \\ \hline
\end{tabular}}
\begin{minipage}{8.5 cm}
	\scriptsize Note. $^*$Best results, $^\phi$Second best results.
\end{minipage}
\end{table}

The second comparison in the analysis (see Figure \ref{fig1}) is designed to evaluate the performance sensitivity of machine learning algorithms with respect to different input variables. In Table \ref{tab11}, XGB refers to the results obtained using principal components derived from climate indexes, while XGB-C corresponds to the scenario where the climate indexes are directly used as input variables. Notably, with the XGB-C method, Model 21, which constitutes the largest data set, consistently yields the best performance, with only a slight difference compared to the second-best models. Among the remaining models that exclude one or more components, the second-best performing models remain consistent with those from the first comparison analysis (Model 19 for corn and wheat, Model 17 for soybeans). One possible explanation for this is that the direct use of climate indexes preserves all original information in the data sets, enabling machine learning algorithms to process and use a greater number of features more effectively, compared to the more compressed data sets derived through principal component analysis. Regional performance metrics of XGB-C are also given in Table \ref{tab102} of the Appendix.

\subsection{Financial Market Case} \label{4.3.}

The analysis of weather derivatives pricing begins by examining the dependence among the three groups (see Table \ref{tab1}). Table \ref{tab3} presents the correlation coefficients ($\rho$) between the weather-based indexes and their corresponding ACI components within each group, along with the $p$-values from the A-D test, which evaluates the null hypothesis that the two distributions are identical. Despite methodological differences in constructing the indexes, strong positive correlations are observed between the corresponding indexes. For the A-D test, $p$-values greater than 0.05 (bold in Table \ref{tab3}) indicate that the paired indexes exhibit statistically similar distributions. The results show that the precipitation-related indexes in Group-3 tend to display similar distributions across most regions compared to temperature-based indexes analyzed in the first two groups. Detailed descriptive statistics for each group are provided in Table \ref{tab4} of the Appendix.

\begin{table}[hbt!]
\centering
\caption{Similarity tests within each group}
\label{tab3}
\scalebox{0.85}{
	\begin{tabular}{|cc|cccccc|}
		\hline
		\multicolumn{2}{|c|}{\multirow{2}{*}{\textbf{}}}              & \multicolumn{6}{c|}{\textbf{Region}}                                                                                                                                                                             \\ \cline{3-8} 
		\multicolumn{2}{|c|}{}                                        & \multicolumn{1}{c|}{\textbf{CEA}}   & \multicolumn{1}{c|}{\textbf{CWP}}    & \multicolumn{1}{c|}{\textbf{MID}}    & \multicolumn{1}{c|}{\textbf{SEA}}    & \multicolumn{1}{c|}{\textbf{SPL}}    & \textbf{SWP}   \\ \hline
		\multicolumn{1}{|c|}{\multirow{2}{*}{\begin{tabular}[c]{@{}c@{}}Group-1\\ (CDD-T90)\end{tabular}}} & $\rho$            & \multicolumn{1}{c|}{0.88}           & \multicolumn{1}{c|}{0.90}            & \multicolumn{1}{c|}{0.89}            & \multicolumn{1}{c|}{0.94}            & \multicolumn{1}{c|}{0.86}            & 0.97           \\ \cline{2-8} 
		\multicolumn{1}{|c|}{}                         & A-D test ($p$-value) & \multicolumn{1}{c|}{0.040} & \multicolumn{1}{c|}{\textless{}0.01} & \multicolumn{1}{c|}{\textless{}0.01} & \multicolumn{1}{c|}{\textless{}0.01} & \multicolumn{1}{c|}{\textless{}0.01} & \textbf{0.250} \\ \hline
		\multicolumn{1}{|c|}{\multirow{2}{*}{\begin{tabular}[c]{@{}c@{}}Group-2\\ (HDD-T10)\end{tabular}}} & $\rho$            & \multicolumn{1}{c|}{0.89}           & \multicolumn{1}{c|}{0.83}            & \multicolumn{1}{c|}{0.86}            & \multicolumn{1}{c|}{0.87}            & \multicolumn{1}{c|}{0.81}            & 0.80           \\ \cline{2-8} 
		\multicolumn{1}{|c|}{}                         & A-D test ($p$-value) & \multicolumn{1}{c|}{0.025}          & \multicolumn{1}{c|}{\textless{}0.01} & \multicolumn{1}{c|}{\textless{}0.01} & \multicolumn{1}{c|}{\textless{}0.01}           & \multicolumn{1}{c|}{\textbf{0.105}}  & \textless{}0.01 \\ \hline
		\multicolumn{1}{|c|}{\multirow{2}{*}{\begin{tabular}[c]{@{}c@{}}Group-3\\ (PRE-P)\end{tabular}}} & $\rho$            & \multicolumn{1}{c|}{0.83}           & \multicolumn{1}{c|}{0.79}            & \multicolumn{1}{c|}{0.88}            & \multicolumn{1}{c|}{0.87}            & \multicolumn{1}{c|}{0.69}            & 0.84           \\ \cline{2-8} 
		\multicolumn{1}{|c|}{}                         & A-D test ($p$-value) & \multicolumn{1}{c|}{\textbf{0.250}} & \multicolumn{1}{c|}{\textbf{0.126}}  & \multicolumn{1}{c|}{\textbf{0.250}}           & \multicolumn{1}{c|}{\textbf{0.189}}  & \multicolumn{1}{c|}{\textless{}0.01}  & \textbf{0.085} \\ \hline
\end{tabular}}
\begin{minipage}{10.4 cm}
	\scriptsize Note. $\rho$: Spearman rank correlation coefficient.
\end{minipage}
\end{table}

For the payoff analysis of derivative options, pairs of indexes with the highest and lowest correlations are selected as case studies. As shown in Table \ref{tab3}, the pair with the strongest correlation is identified in Group-1 for the SWP region ($\rho=0.97$), while the weakest correlation is found in Group-3 for the SPL region ($\rho=0.69$). Call option payoffs are calculated using Equation (\ref{po}), based on the index values from these two groups and regions.

Additionally, the key assumption of the HBA and IM pricing methods, which is that the underlying index should be time-stationary, is also considered. According to the ADF test results (not shown), the CDD and T90 indexes in Group-1 for the SWP region are determined to be non-stationary; therefore, these series are detrended using Equation (\ref{e91}) prior to inclusion in the analyses.

For consistency, the tick size ($\alpha$) in the Equation (\ref{po}) is set to 100 for both groups, and the strike level ($K$) is defined as the mean of the historical index values plus 20\% of their standard deviation \citep{r1}. In the HBA method, payoffs for each year are computed, and the correlations between the payoffs and their A-D test results are analyzed. For the IM method, the best-fitting distributions for the historical indexes are identified, and 1,000 index values are simulated from these distributions to compute payoffs. Finally, the A-D test is applied to the simulated payoffs values. The results of the similarity tests comparing the payoffs obtained from both methods are summarized in Table \ref{tab38}.

\begin{table}[hbt!]
\centering
\caption{Similarity tests between payoffs for each method}
\label{tab38}
\scalebox{0.8}{
	\begin{tabular}{|c|c|lc|cc|cc|}
		\hline
		\multirow{2}{*}{\textbf{Method}} & \multirow{2}{*}{\textbf{N}} & \multicolumn{2}{c|}{\multirow{2}{*}{\textbf{Compared Parameters}}} & \multicolumn{2}{c|}{\textbf{Group-1 (SWP)}}      & \multicolumn{2}{c|}{\textbf{Group-3 (SPL)}}    \\ \cline{5-8} 
		&                             & \multicolumn{2}{c|}{}                                              & \multicolumn{1}{c|}{\textbf{CDD}} & \textbf{T90} & \multicolumn{1}{c|}{\textbf{PRE}} & \textbf{P} \\ \hline
		\multirow{2}{*}{HBA}             & \multirow{2}{*}{64}         & \multicolumn{1}{l|}{\multirow{2}{*}{Payoffs}} & $\rho$                  & \multicolumn{2}{c|}{0.82}                        & \multicolumn{2}{c|}{0.65}                      \\ \cline{4-8} 
		&                             & \multicolumn{1}{l|}{}                         & A-D test ($p$-value)       & \multicolumn{2}{c|}{0.250}                       & \multicolumn{2}{c|}{0.250}                     \\ \hline
		\multirow{2}{*}{IM}              & 64                          & \multicolumn{1}{l|}{Indexes}                  & Best-fitting dist. & \multicolumn{1}{c|}{weibull}      & beta         & \multicolumn{1}{c|}{lognorm}      & weibull    \\ \cline{2-8} 
		& 1,000                       & \multicolumn{1}{l|}{Payoffs}                  & A-D test ($p$-value)       & \multicolumn{2}{c|}{\textless{}0.01}                       & \multicolumn{2}{c|}{0.038}                     \\ \hline
\end{tabular}}
\begin{minipage}{10 cm}
	\hspace{-0.2cm}
	\scriptsize Note. $\rho$: Spearman rank correlation coefficient, N: Number of years.
\end{minipage}
\end{table}

According to Table \ref{tab38}, the correlations among the payoffs obtained using the HBA method closely align with the correlations observed between the original indexes. For Group-1 (SWP), the correlation coefficient between the payoffs is 0.82 (down from 0.97), while for Group-3 (SPL), it is 0.65 (decreasing from 0.69). The decline observed for Group-1 (SWP) is expected, as the CDD and T90 indexes are detrended prior to the analysis. Moreover, based on the A-D test results for the HBA method ($p > 0.05$), the payoff distributions are statistically similar. Interestingly, although the A-D test results for Group-3 (SPL) in Table \ref{tab3} show no distributional similarity between the indexes themselves, the observed similarity among their payoffs implies that these indexes exhibit consistent behavior in terms of financial outcomes, highlighting their potential applicability in weather-derivative markets.

Table \ref{tab38} reports the best-fitting distributions for each index pair (with detrended CDD and T90). Distinct best-fitting distributions are identified for all indexes, a finding further confirmed by the A-D test results derived from 1,000 simulated indexes and their corresponding payoffs ($p < 0.05$). In contrast to the HBA findings, no similar distributions are detected among the simulated payoffs in the IM approach, indicating that the degree of similarity among payoffs in simulation-based analyses strongly depends on the distributional characteristics of the underlying indexes themselves.

\begin{figure}[htb!]
\centering
\begin{subfigure}[b]{0.45\textwidth}
	\centering
	\includegraphics[scale=0.45]{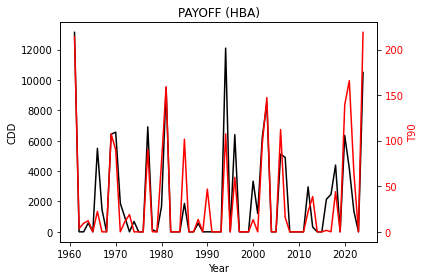}
	\includegraphics[scale=0.45]{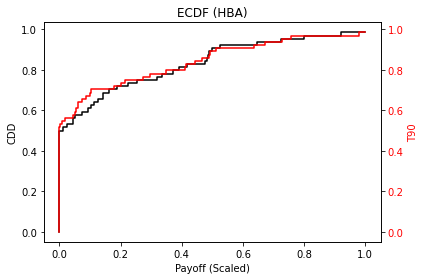}
	\includegraphics[scale=0.45]{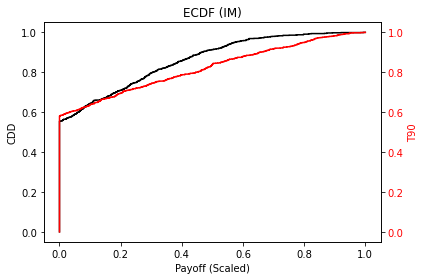}
	\caption{Group-1 (SWP)}
\end{subfigure}
\hspace{1cm}
\begin{subfigure}[b]{0.45\textwidth}
	\centering
	\includegraphics[scale=0.45]{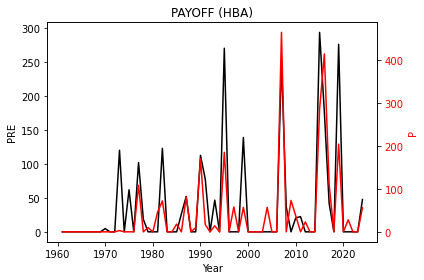}
	\includegraphics[scale=0.45]{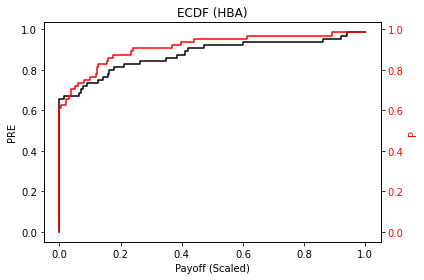}
	\includegraphics[scale=0.45]{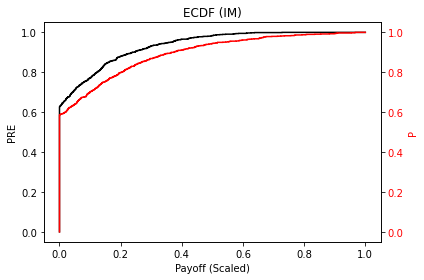}
	\caption{Group-3 (SPL)}
\end{subfigure}
\caption{Payoffs and ECDF values for each method}
\label{fig31}
\end{figure}

The payoffs ($PO$) and ECDF values derived from both methods are illustrated in Figure \ref{fig31}. Examination of the payoff graphs reveals that the call options constructed from the indexes within each group often generate payments during similar time periods. This pattern indicates that the indexes tend to reach extreme values simultaneously, confirming their consistency and suitability for use in weather-derivative markets.

The payoffs ($PO$) of uncapped call options are computed based on Equation (\ref{po}). The fair prices are then obtained by averaging these payoffs using the HBA method. For Group-1, the fair price is calculated as 2,192.23 for CDD and 34.74 for T90. Similarly, for Group-3, the corresponding fair prices are 36.34 for PRE and 40.61 for P. The observed discrepancies in fair prices arise from the fact that each index expresses weather conditions using different measurement units. Recall that CDD quantifies temperature in degree-days, whereas PRE measures precipitation in millimeters. In contrast, the ACI components T90 and P express weather conditions as standardized anomalies relative to a reference period. Comparing the prices obtained using T90 and P, which share identical units, indicates that the call option priced with P is more expensive than that based on T90 (40.61 vs. 34.74). This finding suggests that extreme precipitation events have historically occurred more frequently than extreme temperature events.

\section{Discussion} \label{sec44}

The evaluation results suggest that the ACI provides a useful and practically relevant framework for representing climate-related conditions, while also highlighting important differences between the informational content of individual ACI components and that of the composite index.

From the agricultural perspective, the findings show that both ACI components and traditional WBIs possess significant predictive power for crop yields. By fitting 22 distinct models, the analysis indicates that the use of multiple climate indicators, rather than reliance on a single aggregate measure, provides a richer representation of the weather conditions relevant to agricultural production. In particular, the inclusion of indexes reflecting additional weather variables-such as wind speed, drought, and sea-level changes-improves predictive performance beyond models based only on temperature and precipitation-related measures. This finding highlights that crop yield dynamics are influenced by a broader set of climate factors than those captured by conventional weather indicators alone.

At the same time, while the individual ACI components contain meaningful predictive information, the composite index alone does not appear to fully capture the multidimensional and crop-specific nature of the weather conditions affecting agricultural production. This implies that aggregation, although convenient, may mask relevant heterogeneity across different climate dimensions. Consequently, for forecasting and risk assessment purposes, using individual components may provide more informative and accurate results than relying solely on the composite index.

The comparison between ACI components and their WBI counterparts also provides important conceptual insights. Despite differences in their construction, the empirical results indicate that these indexes often provide similar information about underlying weather conditions. This supports the broader motivation of the study, namely that the ACI can serve as a viable alternative representation of climate conditions in contexts where traditional weather indexes have been widely used.

From the financial market perspective, the results show that ACI components and their WBI counterparts can generate similar financial behavior. The strong dependence observed among the payoff structures suggests that the ACI framework can be effectively incorporated into weather-derivative applications, including hedging and risk-transfer strategies. Notably, this extends the use of the ACI beyond descriptive climate analysis into a market-based pricing framework, reinforcing its applicability in real-world financial contexts.

The findings have important practical implications for decision-making. In agriculture, the use of multiple climate indicators can improve crop yield forecasting and support decisions related to planting, resource allocation, and risk management. In financial markets, the consistency observed in derivative pricing suggests that ACI-based variables can be effectively incorporated into hedging and risk-transfer strategies. Thus, the components of the ACI can serve as practical decision-support tools by transforming complex climate information into operational and interpretable metrics.

Nevertheless, several limitations should be acknowledged. First, the crop yield analysis is based on regional averages, which may not fully reflect heterogeneity across individual states or smaller spatial units. Second, the empirical analysis is conducted using U.S. data and a specific set of crops, which may limit the generalizability of the findings to other regions or agricultural systems. Third, although the weather-derivative analysis demonstrates the usefulness of ACI components in a pricing context, it is based on a specific contract structure and selected pricing methods. Broader financial applications may require alternative derivative designs or more elaborate stochastic modeling frameworks.

In addition to these findings, the proposed methodology contributes to the literature on crop yield forecasting. Forecasting weather-related factors that significantly affect crop yields such as temperature, precipitation, and sea-level, often requires complex, time-consuming methods due to their inherently dynamic nature. However, by incorporating future projections of components derived through the FPCA method, the proposed framework enables more efficient and practical crop yield forecasting.

\section{Conclusion} \label{sec5}

This study examines the Actuaries Climate Index\textsuperscript{TM} in both agricultural and financial applications by comparing it with traditional weather-based indexes. Across crop yield prediction and weather-derivative pricing, the findings demonstrate that the ACI framework provides a meaningful and practically relevant representation of climate-related conditions.

In the agricultural application, the results indicate that both ACI components and traditional WBIs exhibit strong predictive capability for crop yields. In the financial application, the analysis shows that ACI components and their WBI counterparts can produce similar derivative payoffs and pricing behavior when they reflect comparable climate conditions.

Overall, these findings highlight the potential of the ACI as a versatile climate risk indicator across different domains. The study contributes to the growing literature on climate-related decision tools and provides a foundation for future research on more targeted, region-specific, and forward-looking applications.

\section*{Acknowledgments}

Jose Garrido gratefully acknowledges the financial support of the Natural Sciences and Engineering Research Council of Canada (RGPIN-2017-06643).

\bibliographystyle{plainnat}
\bibliography{kaynaklar}

\clearpage

\appendix

\section*{Appendix} \label{bbb1}

\setcounter{table}{0}
\setcounter{figure}{0}
\renewcommand{\thetable}{A\arabic{table}}
\renewcommand{\thefigure}{A\arabic{figure}}

\begin{table}[hbt!]
	\centering
	\caption{Detailed structure of explanatory variables included in each model}
	\label{tab22}
	\begin{adjustwidth}{1cm}{0cm}
		\scalebox{0.76}{
\begin{tabular}{|c|cccccccccccccccccccccc|}
				\hline
				\multirow{2}{*}{\textbf{Variable}} & \multicolumn{22}{c|}{\textbf{Model}}                                                                                                                                                                                                                                                                                                                                                                                                                                                                                                                                                                                                                                                                                                                              \\ \cline{2-23} 
				& \multicolumn{1}{c|}{\textbf{1}} & \multicolumn{1}{c|}{\textbf{2}} & \multicolumn{1}{c|}{\textbf{3}} & \multicolumn{1}{c|}{\textbf{4}} & \multicolumn{1}{c|}{\textbf{5}} & \multicolumn{1}{c|}{\textbf{6}} & \multicolumn{1}{c|}{\textbf{7}} & \multicolumn{1}{c|}{\textbf{8}} & \multicolumn{1}{c|}{\textbf{9}} & \multicolumn{1}{c|}{\textbf{10}} & \multicolumn{1}{c|}{\textbf{11}} & \multicolumn{1}{c|}{\textbf{12}} & \multicolumn{1}{c|}{\textbf{13}} & \multicolumn{1}{c|}{\textbf{14}} & \multicolumn{1}{c|}{\textbf{15}} & \multicolumn{1}{c|}{\textbf{16}} & \multicolumn{1}{c|}{\textbf{17}} & \multicolumn{1}{c|}{\textbf{18}} & \multicolumn{1}{c|}{\textbf{19}} & \multicolumn{1}{c|}{\textbf{20}} & \multicolumn{1}{c|}{\textbf{21}} & \textbf{22} \\
\hline
		\end{tabular}}
	\end{adjustwidth}
\end{table}

\clearpage

\begin{center}
	
	\rotatebox{90}{
		\begin{minipage}{0.75\textheight}
			\centering
			\parbox{10cm}{
				\captionof{table}[Short Title]{Descriptive statistics of each group \label{tab4}}
			}
			\resizebox{!}{\height}{%
				\scalebox{0.9}{
					\begin{tabular}{|ccc|cccccc|}
						\hline
						\multicolumn{3}{|c|}{\multirow{2}{*}{\textbf{N = 64}}}                                                 & \multicolumn{6}{c|}{\textbf{Region}}                                                                                                                                                             \\ \cline{4-9} 
						\multicolumn{3}{|c|}{}                                                                                 & \multicolumn{1}{c|}{\textbf{CEA}} & \multicolumn{1}{c|}{\textbf{CWP}} & \multicolumn{1}{c|}{\textbf{MID}} & \multicolumn{1}{c|}{\textbf{SEA}} & \multicolumn{1}{c|}{\textbf{SPL}} & \textbf{SWP} \\ \hline
						\multicolumn{1}{|c|}{\multirow{6}{*}{Group-1}} & \multicolumn{1}{c|}{\multirow{3}{*}{CDD}} & Mean \text{\textpm } SD & \multicolumn{1}{c|}{507 \text{\textpm } 98}     & \multicolumn{1}{c|}{245 \text{\textpm } 73}     & \multicolumn{1}{c|}{618 \text{\textpm } 109}    & \multicolumn{1}{c|}{1178 \text{\textpm } 110}   & \multicolumn{1}{c|}{795 \text{\textpm } 90}     & 892 \text{\textpm } 105    \\ \cline{3-9} 
						\multicolumn{1}{|c|}{}                         & \multicolumn{1}{c|}{}                     & Min.      & \multicolumn{1}{c|}{326}          & \multicolumn{1}{c|}{85}           & \multicolumn{1}{c|}{356}          & \multicolumn{1}{c|}{929}          & \multicolumn{1}{c|}{527}          & 697          \\ \cline{3-9} 
						\multicolumn{1}{|c|}{}                         & \multicolumn{1}{c|}{}                     & Max.      & \multicolumn{1}{c|}{704}          & \multicolumn{1}{c|}{428}          & \multicolumn{1}{c|}{839}          & \multicolumn{1}{c|}{1465}         & \multicolumn{1}{c|}{1017}         & 1145         \\
\hline
				\end{tabular}}
				\begin{minipage}{18.5 cm}
					\vspace{7.8cm}
					\hspace{-16.1cm}
					\footnotesize Note. N: Number of observations (years), SD: Standard deviation, Min: Minimum, Max: Maximum.
				\end{minipage}
			}%
		\end{minipage}
	}
\end{center}

\clearpage

\begin{sidewaystable}[hbt!]
	\centering
	\caption{MAPE values of GLM for each crop, model, and region}\label{tab6}
	\scalebox{0.72}{
		\begin{tabular}{|c|c|c|cccccccccccccccccccccc|}
			\hline
			\multirow{2}{*}{\textbf{Set}} & \multirow{2}{*}{\textbf{Crop}} & \multirow{2}{*}{\textbf{Region}} & \multicolumn{22}{c|}{\textbf{Model}}                                                                                                                                                                                                                                                                                                                                                                                                                                                                                                                                                                                                                                                                                                                              \\ \cline{4-25} 
			&                                &                                  & \multicolumn{1}{c|}{\textbf{1}} & \multicolumn{1}{c|}{\textbf{2}} & \multicolumn{1}{c|}{\textbf{3}} & \multicolumn{1}{c|}{\textbf{4}} & \multicolumn{1}{c|}{\textbf{5}} & \multicolumn{1}{c|}{\textbf{6}} & \multicolumn{1}{c|}{\textbf{7}} & \multicolumn{1}{c|}{\textbf{8}} & \multicolumn{1}{c|}{\textbf{9}} & \multicolumn{1}{c|}{\textbf{10}} & \multicolumn{1}{c|}{\textbf{11}} & \multicolumn{1}{c|}{\textbf{12}} & \multicolumn{1}{c|}{\textbf{13}} & \multicolumn{1}{c|}{\textbf{14}} & \multicolumn{1}{c|}{\textbf{15}} & \multicolumn{1}{c|}{\textbf{16}} & \multicolumn{1}{c|}{\textbf{17}} & \multicolumn{1}{c|}{\textbf{18}} & \multicolumn{1}{c|}{\textbf{19}} & \multicolumn{1}{c|}{\textbf{20}} & \multicolumn{1}{c|}{\textbf{21}} & \textbf{22} \\
\hline
	\end{tabular}}
	
\end{sidewaystable}

\clearpage

\begin{sidewaystable}[hbt!]
	\centering
	\caption{MAPE values of GAM for each crop, model, and region}\label{tab8}
	\scalebox{0.72}{
		\begin{tabular}{|c|c|c|cccccccccccccccccccccc|}
			\hline
			\multirow{2}{*}{\textbf{Set}} & \multirow{2}{*}{\textbf{Crop}} & \multirow{2}{*}{\textbf{Region}} & \multicolumn{22}{c|}{\textbf{Model}}                                                                                                                                                                                                                                                                                                                                                                                                                                                                                                                                                                                                                                                                                                                              \\ \cline{4-25} 
			&                                &                                  & \multicolumn{1}{c|}{\textbf{1}} & \multicolumn{1}{c|}{\textbf{2}} & \multicolumn{1}{c|}{\textbf{3}} & \multicolumn{1}{c|}{\textbf{4}} & \multicolumn{1}{c|}{\textbf{5}} & \multicolumn{1}{c|}{\textbf{6}} & \multicolumn{1}{c|}{\textbf{7}} & \multicolumn{1}{c|}{\textbf{8}} & \multicolumn{1}{c|}{\textbf{9}} & \multicolumn{1}{c|}{\textbf{10}} & \multicolumn{1}{c|}{\textbf{11}} & \multicolumn{1}{c|}{\textbf{12}} & \multicolumn{1}{c|}{\textbf{13}} & \multicolumn{1}{c|}{\textbf{14}} & \multicolumn{1}{c|}{\textbf{15}} & \multicolumn{1}{c|}{\textbf{16}} & \multicolumn{1}{c|}{\textbf{17}} & \multicolumn{1}{c|}{\textbf{18}} & \multicolumn{1}{c|}{\textbf{19}} & \multicolumn{1}{c|}{\textbf{20}} & \multicolumn{1}{c|}{\textbf{21}} & \textbf{22} \\
\hline
	\end{tabular}}
	
\end{sidewaystable}

\clearpage

\begin{sidewaystable}[hbt!]
	\centering
	\caption{MAPE values of XGB for each crop, model, and region}\label{tab81}
	\scalebox{0.72}{
		\begin{tabular}{|c|c|c|cccccccccccccccccccccc|}
			\hline
			\multirow{2}{*}{\textbf{Set}} & \multirow{2}{*}{\textbf{Crop}} & \multirow{2}{*}{\textbf{Region}} & \multicolumn{22}{c|}{\textbf{Model}}                                                                                                                                                                                                                                                                                                                                                                                                                                                                                                                                                                                                                                                                                                                              \\ \cline{4-25} 
			&                                &                                  & \multicolumn{1}{c|}{\textbf{1}} & \multicolumn{1}{c|}{\textbf{2}} & \multicolumn{1}{c|}{\textbf{3}} & \multicolumn{1}{c|}{\textbf{4}} & \multicolumn{1}{c|}{\textbf{5}} & \multicolumn{1}{c|}{\textbf{6}} & \multicolumn{1}{c|}{\textbf{7}} & \multicolumn{1}{c|}{\textbf{8}} & \multicolumn{1}{c|}{\textbf{9}} & \multicolumn{1}{c|}{\textbf{10}} & \multicolumn{1}{c|}{\textbf{11}} & \multicolumn{1}{c|}{\textbf{12}} & \multicolumn{1}{c|}{\textbf{13}} & \multicolumn{1}{c|}{\textbf{14}} & \multicolumn{1}{c|}{\textbf{15}} & \multicolumn{1}{c|}{\textbf{16}} & \multicolumn{1}{c|}{\textbf{17}} & \multicolumn{1}{c|}{\textbf{18}} & \multicolumn{1}{c|}{\textbf{19}} & \multicolumn{1}{c|}{\textbf{20}} & \multicolumn{1}{c|}{\textbf{21}} & \textbf{22} \\
\hline
	\end{tabular}}
	
\end{sidewaystable}

\clearpage

\begin{sidewaystable}[hbt!]
	\centering
	\caption{MAPE values of LGBM for each crop, model, and region}\label{tab82}
	\scalebox{0.72}{
		\begin{tabular}{|c|c|c|cccccccccccccccccccccc|}
			\hline
			\multirow{2}{*}{\textbf{Set}} & \multirow{2}{*}{\textbf{Crop}} & \multirow{2}{*}{\textbf{Region}} & \multicolumn{22}{c|}{\textbf{Model}}                                                                                                                                                                                                                                                                                                                                                                                                                                                                                                                                                                                                                                                                                                                              \\ \cline{4-25} 
			&                                &                                  & \multicolumn{1}{c|}{\textbf{1}} & \multicolumn{1}{c|}{\textbf{2}} & \multicolumn{1}{c|}{\textbf{3}} & \multicolumn{1}{c|}{\textbf{4}} & \multicolumn{1}{c|}{\textbf{5}} & \multicolumn{1}{c|}{\textbf{6}} & \multicolumn{1}{c|}{\textbf{7}} & \multicolumn{1}{c|}{\textbf{8}} & \multicolumn{1}{c|}{\textbf{9}} & \multicolumn{1}{c|}{\textbf{10}} & \multicolumn{1}{c|}{\textbf{11}} & \multicolumn{1}{c|}{\textbf{12}} & \multicolumn{1}{c|}{\textbf{13}} & \multicolumn{1}{c|}{\textbf{14}} & \multicolumn{1}{c|}{\textbf{15}} & \multicolumn{1}{c|}{\textbf{16}} & \multicolumn{1}{c|}{\textbf{17}} & \multicolumn{1}{c|}{\textbf{18}} & \multicolumn{1}{c|}{\textbf{19}} & \multicolumn{1}{c|}{\textbf{20}} & \multicolumn{1}{c|}{\textbf{21}} & \textbf{22} \\
\hline
	\end{tabular}}
	
\end{sidewaystable}

\clearpage

\begin{sidewaystable}[hbt!]
	\centering
	\caption{MAPE values of XGB-C for each crop, model, and region}\label{tab102}
	\scalebox{0.72}{
		\begin{tabular}{|c|c|c|cccccccccccccccccccccc|}
			\hline
			\multirow{2}{*}{\textbf{Set}} & \multirow{2}{*}{\textbf{Crop}} & \multirow{2}{*}{\textbf{Region}} & \multicolumn{22}{c|}{\textbf{Model}}                                                                                                                                                                                                                                                                                                                                                                                                                                                                                                                                                                                                                                                                                                                              \\ \cline{4-25} 
			&                                &                                  & \multicolumn{1}{c|}{\textbf{1}} & \multicolumn{1}{c|}{\textbf{2}} & \multicolumn{1}{c|}{\textbf{3}} & \multicolumn{1}{c|}{\textbf{4}} & \multicolumn{1}{c|}{\textbf{5}} & \multicolumn{1}{c|}{\textbf{6}} & \multicolumn{1}{c|}{\textbf{7}} & \multicolumn{1}{c|}{\textbf{8}} & \multicolumn{1}{c|}{\textbf{9}} & \multicolumn{1}{c|}{\textbf{10}} & \multicolumn{1}{c|}{\textbf{11}} & \multicolumn{1}{c|}{\textbf{12}} & \multicolumn{1}{c|}{\textbf{13}} & \multicolumn{1}{c|}{\textbf{14}} & \multicolumn{1}{c|}{\textbf{15}} & \multicolumn{1}{c|}{\textbf{16}} & \multicolumn{1}{c|}{\textbf{17}} & \multicolumn{1}{c|}{\textbf{18}} & \multicolumn{1}{c|}{\textbf{19}} & \multicolumn{1}{c|}{\textbf{20}} & \multicolumn{1}{c|}{\textbf{21}} & \textbf{22} \\
\hline
	\end{tabular}}
	
\end{sidewaystable}

\end{document}